\documentclass[12pt]{article}
\usepackage{amsmath,cite}
\usepackage{amssymb}
\newcommand{\p}[1]{(\ref{#1})}
\newcommand{\be}{\begin{equation}}
\newcommand{\ee}{\end{equation}}
\newcommand{\bea}{\begin{eqnarray}}
\newcommand{\eea}{\end{eqnarray}}
\def\theequation{\arabic{section}.\arabic{equation}}
\def\ba{\begin{array}} \def\ea{\end{array}}

\begin{document}
\topmargin -1cm \oddsidemargin=0.25cm\evensidemargin=0.25cm
\setcounter{page}0
\renewcommand{\thefootnote}{\fnsymbol{footnote}}
\begin{titlepage}
\vskip .7in
\begin{center}
{\Large \bf  Higher Spin Fields in Hyperspace. A Review
 } \vskip .7in 
 {\Large 
Dmitri Sorokin$^a$\footnote{e-mail: {\tt  dmitri.sorokin@pd.infn.it }} and
 Mirian Tsulaia$^b$\footnote{e-mail: {\tt  mirian.tsulaia@uwa.edu.au}  }}
 \vskip .4in {$^a$ \it INFN, Sezione di Padova, via F. Marzolo 8, 35131 Padova, Italia} \\
\vskip .2in { $^b$ \it School of Physics $M013$,
The University of
Western Australia, 35 Stirling Highway,
Crawley, Perth, WA 6009, Australia}\\
\vskip .8in
\begin{abstract}

We give an introduction to the so-called tensorial, matrix or
hyperspace approach to the description of massless higher-spin fields.

\end{abstract}

\end{center}

\vfill

\end{titlepage}

\tableofcontents

\renewcommand{\thefootnote}{\arabic{footnote}}

\section{Introduction}

Every consistent theory of interacting higher spin fields necessarily includes
an infinite number of such fields. For this reason, it is extremely important to develop a formalism which effectively includes an infinite number of fields into a simpler field-theoretical object.
This formalism should yield correct field equations first of all at the free level and then be promoted to an interacting theory. An elegant geometrical approach to higher spin theories of this kind is known as the method of tensorial spaces. This approach
was first suggested by Fronsdal \cite{Fronsdal:1985pd}. Its explicit dynamical realization and further extensive developments have been carried out in \cite{Bandos:1998vz,Bandos:1999qf,Vasiliev:2001zy,Vasiliev:2001dc,Vasiliev:2002fs,Didenko:2003aa,Plyushchay:2003gv,Gelfond:2003vh,Plyushchay:2003tj,Vasiliev:2003jc,Bandos:2004nn,Bandos:2005mb,Ivanov:2005ss,Gelfond:2006be,Vasiliev:2007yc,Ivanov:2007vx,West:2007mh,Gelfond:2008ur,Gelfond:2008td,Gelfond:2010pm,Bandos:2011wi,Florakis:2014kfa,Florakis:2014aaa,Fedoruk:2012ka,Gelfond:2015poa,Skvortsov:2016lbh,Goncharov:2016wxw}.

In a certain sense, the
method of tensorial spaces is reminiscent of the Kaluza-Klein theories. In such theories, one
usually considers massless field equations in higher dimensions and then, assuming that the
extra dimensions are periodic (compact), one obtains a theory in lower dimensions, which contains fields with growing masses. In the method of tensorial (super)spaces, one also considers
theories in multi-dimensional space-times, but in this case the extra dimensions are introduced
in such a way that they generate the fields with higher spins instead of the fields with increasing masses.
A main advantage of the formulation of the higher spin theories on extended tensorial
(super)spaces is that one can combine curvatures of an infinite number of bosonic and fermionic
higher spin fields into a single ``master" (or ``hyper") scalar and spinor field which propagate
through the tensorial supesrpaces (also called hyperspaces). The field equations in the tensorial
spaces are invariant under the action of $Sp(2n)$ group, whereas the dimensions of the corresponding tensorial spaces are equal to $\frac{n(n+1)}2$.
The case of four space-time dimensions $D = 4$ is of
particular interest since the approach of tensorial (super)spaces comprises all massless higher
spin fields from zero to infinity. The free field equations are invariant under the $Sp(8)$ group,
which contains a four dimensional conformal group $SO(2,4)$ as a subgroup. In fact, the entire
structure of the $Sp(8)$ invariant formulation of the higher spin fields is a straightforward generalization of the conformally invariant formulation of the four-dimensional scalar and spinor
fields. This allows one to use the experience and intuition gained from the usual conformal field
theories for studying the dynamics of higher spin fields on flat and AdS backgrounds, and to construct their correlation functions.

Being intrinsically related to the unfolded formulation \cite{Vasiliev:1988xc,Vasiliev:1988sa,Vasiliev:2001ur,Vasiliev:2004qz,Vasiliev:2012vf} of higher-spin field theory, the hyperspace approach provides an extra and potentially powerful tool for studying
higher spin AdS/CFT correspondence (for reviews on higher-spin holography, see, e.g. \cite{Giombi:2012ms,Gaberdiel:2012uj}). The origin of higher-spin holographic duality can be
traced back to the work of Flato and Fronsdal \cite{Flato:1978qz} who showed that the tensor product of
single-particle states of a 3D massless conformal scalar and spinor fields (singletons) produces
the tower of all single-particle representations of 4D massless fields whose spectrum matches
that of 4D higher spin gauge theories. The hyperspace formulation provides an explicit field
theoretical realization of the Flato-Fronsdal theorem in which higher spin fields are embedded
in a single scalar and spinor fields, though propagating in hyperspace. The relevance of the
unfolded and hyperspace formulation to the origin of holography has been pointed out in \cite{Vasiliev:2012vf}. In this interpretation,
holographically dual theories share the same unfolded formulation in extended spaces which
contains twistor-like variables and each of these theories corresponds to a different
reduction, or ``visualization", of the same ``master" theory.

In what follows, we will review main features and latest developments of the tensorial space approach, and associated generalized conformal theories. It is  mainly based on the papers \cite{Bandos:1999qf,Plyushchay:2003gv,Plyushchay:2003tj,Bandos:2005mb,Florakis:2014kfa,Florakis:2014aaa,Skvortsov:2016lbh}. We hope that this will be a useful complement to a number of available reviews on the higher-spin gauge theories which reflect other aspects and different approaches to the subject
\begin{itemize}
\item Frame-like approach in higher-spin field theory \cite{Vasiliev:1995dn,Vasiliev:1999ba,Bekaert:2005vh,Sezgin:2012ag,Didenko:2014dwa,Arias:2016ajh}.
\item Metric-like approach \cite{Bekaert:2003uc, Bouatta:2004kk,Francia:2006hp, Fotopoulos:2008ka,Campoleoni:2009je,Francia:2010ap,Bekaert:2010hw,Taronna:2010qq, Sagnotti:2013bha,Joung:2012fv,Taronna:2012gb,Gomez:2014dwa,Leonard:2017uvo}.
\item Review that address the both approaches \cite{Rahman:2015pzl}.
\item Higher-spin Holography   \cite{Gaberdiel:2012uj,Giombi:2012ms,  Sleight:2016hyl,Sleight:2017krf}.
\item Reviews which contain both the metric-like approach and the hyperspace approach \cite{Sorokin:2004ie,Tsulaia:2012rb}.
\item A short review on the hyperspace approach \cite{Bandos:2005rr}.
\item A short review that contains frame-like approach, hyperspaces and higher-spin holography \cite{Vasiliev:2014vwa}.
\end{itemize}

The review is organized as follows. In  Section \ref{FH} we introduce a general concept of flat hyperspaces.
To this end we use somewhat heuristic argument, which includes a direct generalization
of the famous twistor-like representation of a light-light momentum of a particle
to higher dimensional tensorial spaces i.e. to hyperspaces.  The basic fields in this set up are
one bosonic and one fermionic hyperfield, which contain infinite sets of bosonic and fermionic field strengths of massless fields with spins ranging from zero to infinity. Physically interesting examples are hyperspaces associated with ordinary space-times of dimensions
$D=3,4,6$ and $10$. In what follows, we will always keep in mind these physical cases, though from the geometric perspective the tensorial spaces of any dimension have the same properties.

We demonstrate in detail that the solutions of  wave equations in hyperspace are generating functionals for
higher spin fields. These equations are nothing but a set of free conformal higher spin equations
in $D=3,4,6$ and $10$. The case of $D=3$ describes only scalar and spinor fields, the case of $D=4$ comprises the all massless bosonic and fermionic higher spin fields with spins
from $0$ to $\infty$ and the cases of $D=6$ and $D=10$ describe infinite sets of fields whose field strengths are self-dual multiforms. These fields carry unitary irreducible representations of the higher-dimensional conformal group and are sometimes called "spinning singletons" \cite{Angelopoulos:1997ij}.

We then describe a generalized conformal group $Sp(2n)$ which contains a convention conformal group $SO(2,D)$ as its subgroup (for $D=3,4,6,10$ and $n=2,4,8,16$, respectively) and show how the coordinates in hyperspace and the hyperfields transform under these generalized conformal transformations.

In  Section \ref{SectionAdS}, we consider an example of curved hyperspaces which are
$Sp(n)$ group manifolds. An interesting property of these manifolds is that
they are hyperspace generalizations of $AdS_D$ spaces.
Similarly to the $AdS_D$ space  which can be regarded as a coset space of the conformal group $SO(2,D)$, the $Sp(n)$ group manifold is a coset space of the generalized conformal group $Sp(2n)$. This results in the fact that
the property of the conformal flatness of the $AdS_D$ spaces (i.e., the existence of a basis in which the $AdS$ metric is proportional to a flat metric) is also generalized to the case of hyperspaces. In particular,
a metric on the $Sp(n)$ group manifold
is flat up to a rotation of the $GL(n)$ group, the property that we call  ``$GL$--flatness".

In Section \ref{Sectionpareticle}, we briefly discuss  how the field equations given in the previous Sections
can be obtained as a result of the  quantization of (super)particle models on hyperspaces.

In Section \ref{sectionfesp}, we  derive the field equations on $Sp(n)$ group manifolds. We show that
the field equations on flat hyperspaces and $Sp(n)$ group manifolds can be transformed into each other by performing
a generalized conformal rescaling of the hyperfields. We discus plane wave solutions on generalized $AdS$ spaces
and present a generalized conformal (i.e. $Sp(2n)$) transformations of the hyperfields
on the $Sp(n)$ group manifolds. In all these considerations, the property of  $GL(n)$ flatness plays a crucial role.

Section \ref{SECTIONSUSY} describes a supersymmetric generalization  of the construction considered in Section \ref{FH} and
 Section \ref{HEADS} deals with the supersymmetric generalization of the field theory on $Sp(n)$ introduced in Section \ref{SectionAdS}.
 The generalization is straightforward but nontrivial.
 Instead of hyperspace, we consider hyper-superspaces and instead of hyperfields we consider
 hyper-superfields.
 The generalized superconformal symmetry is
  the $OSp(1|2n)$ supergroup and the generalized super-$AdS$ spaces are $OSp(1|n)$ supergroup manifolds.
 We show that all the characteristic features of the hyperspaces and hyperfield equations are generalized to the supersymmetric case as well.

The direct analogy with usual $D$-dimensional CFTs suggests a possibility of considering
generalized conformal field theories in hyperspaces.
Sections \ref{s5} and \ref{sections6} deal  with such a theory which is based
on the invariance of correlation functions under the generalized
conformal group $Sp(2n)$. The technique used in these Sections
is borrowed  from usual $D$-dimensional CFTs and the correlation functions are obtained via solving the generalized Ward identities in (super) hyperspaces.

In Section \ref{s5}, we derive $OSp(1|2n)$ invariant two--, three-- and four-point functions for
scalar super-hyperfields. The correlation functions
for component fields can be obtained by simply expanding the results in series of the powers of Garssmann
coordinates. Therefore, we shall not consider the derivation of $Sp(2n)$ invariant correlation functions for the component fields separately.

Finally, in  Section \ref{sections6}, we introduce generalized conserved currents and generalized
stress-tensors. Their explicit forms and the transformation rules under $Sp(2n)$ can be readily
obtained from the free field equations and the transformation rules of the free hyperfields.

Further we show how one can compute $Sp(2n)$ invariant correlation functions which involve the basic
hyperfields together with higher rank tensors such as conserved currents
and the generalized  stress tensor.
We show that the $Sp(2n)$ invariance itself does not impose
any restriction on the generalized conformal dimensions of the basic hyperfields even if the  conformal dimensions of the current and stress tensor remains canonical.

However, the further requirements of the conservation of the generalized current and generalized stress
tensor fixes also the conformal dimensions of the basic hyperfields, implying
that the generalized conformal theory will not allow for nontrivial interactions.

We briefly discuss possibilities of avoiding these restrictions by considering spontaneously broken
$Sp(2n)$ symmetry or local $Sp(2n)$ invariance, which may lead to an interacting hyperfield theory.

Appendices contain some technical details such as conventions used in the review,
a
derivation of the field equations on $Sp(n)$ group manifolds and some useful identities.

\section{Flat hyperspace} \label{FH} \setcounter{equation}0

Let us formulate the basic idea behind the introduction of tensorial space.
We shall mainly concentrate on a tensorial extension of four-dimensional
Minkowski space--time. A generalization to higher dimensional $D=6$ and
$D=10$ spaces will be given later in this Section.

~Consider ~a ~four ~dimensional ~massless ~scalar ~field. ~Its ~light--like ~momentum
$p_m p^m=0$, $m=0,1,2,3$ ~can ~be ~expressed via the Cartan-Penrose (twistor) representation as a bilinear combination of a commuting Weyl spinor
$\lambda_A$ and its complex conjugate $\overline \lambda_{\dot A}$ $(A, {\dot A}=1,2)$
\begin{equation} \label{BW}
p^m = \lambda^A (\sigma^m)_{A\dot A}\tilde\lambda^{\dot A},  \quad or \quad
    P_{A \dot A} = \lambda_A  \overline \lambda_{\dot A}.
\end{equation}
Obviously, since the spinors are commuting, one has $\lambda^A\lambda^B\varepsilon_{AB}\equiv\lambda^A \lambda_A=0=\overline \lambda^{\dot A} \overline \lambda_{\dot A}$
and therefore $P^{A \dot A}P_{A \dot A}=0$, where the spinor indices are raised and lowered with the unit antisymmetric tensors $\varepsilon^{AB}$ and $\varepsilon_{AB}$.

In order to generalize this construction to higher dimensions note that
one can equivalently rewrite the equation (\ref{BW}) in terms of
four-dimensional real Majorana spinors $\lambda^{\alpha}$ ($\alpha = 1,...,4$)
\begin{equation} \label{BM}
p^m = \lambda^\alpha  \gamma_{\alpha \beta}^m \lambda^{\beta}.
\end{equation}
Due to the Fierz identities
\begin{equation} \label{D101}
(\gamma^m)_{\alpha \beta} (\gamma_m)_{\gamma \delta}+ (\gamma^m)_{\alpha \delta} (\gamma_m)_{\beta \gamma}+
(\gamma^m)_{\alpha \gamma} (\gamma_m)_{\delta \beta} =0\,
\end{equation}
satisfied by  the Dirac matrices $(\gamma^m)_{\alpha \beta}=(\gamma^m)_{ \beta \alpha}$
one has $p^m p_m=0$ \footnote{The four-component spinor indices are raised and lowered by antisymmetric charge conjugation matrices $C^{\alpha\beta}$ and $C_{\alpha\beta}$ see the Appendix
\ref{Appendix A}.}. Let us note that since identities similar to
(\ref{D101}) hold
also
in $D=3, 6$ and  $10$, the Cartan-Penrose relation \eqref{BM} is valid in these dimensions as well.

Let us continue with the four-dimensional case. The momentum
$P_{A \dot A} $ is canonically conjugate to coordinates
$x^{A \dot A}$. One can easily solve the quantum analogue
of the equation (\ref{BW})
\begin{equation} \label{PWE}
\left ( \frac{\partial}{\partial x^{A \dot A}} - i\lambda_A  \overline \lambda_{\dot A} \right ) \Phi(x, \lambda)=0
\end{equation}
to obtain a
plane wave solution for the massless scalar particle
\begin{equation}\label{PWWS}
\Phi(x, \lambda,\bar\lambda)= \phi(\lambda,\bar\lambda) e^{i x^{A \dot A} \lambda_A  \overline \lambda_{\dot A}},
\end{equation}
or in terms of the Majorana spinors
\begin{equation}\label{PWMS}
\Phi(x, \lambda)= \phi(\lambda) e^{i x_{m} \lambda^\alpha \gamma^m_{\alpha \beta}  \ \lambda^{\beta}},
\end{equation}
with $\phi(\lambda)$ being an arbitrary spinor function.

Let us now consider the equation
\begin{equation}\label{PRE}
P_{\alpha \beta} = \lambda_{\alpha} \lambda_\beta,
\end{equation}
which looks like a straightforward generalization of (\ref{BW}) and see its implications.
A space-time  described by the coordinates $X^{\alpha \beta}$ (conjugate to $P_{\alpha \beta}$)
is now ten-dimensional, since $X^{\alpha \beta}$ is a  $4 \times 4$ symmetric matrix.
 A basis of symmetric matrices is formed
by the four Dirac matrices $\gamma^m_{\alpha \beta}$ and their six antisymmetric products $\gamma^{mn}_{\alpha \beta}=-\gamma^{mn}_{\alpha \beta}$.
In this basis, $X^{\alpha \beta}$ has the following expansion
 \begin{equation} \label{tensorial}
 X^{\alpha \beta} = \frac{1}{2} x^m (\gamma_m)^{\alpha \beta} + \frac{1}{4} y^{mn}(\gamma_{mn})^{\alpha \beta}.
 \end{equation}
  The analogue of the wave equation (\ref{PWE})
 is now
 \begin{equation} \label{PRR}
\left ( \frac{\partial}{\partial X^{\alpha \beta}} - i\lambda_\alpha  \lambda_{\beta} \right ) \Phi(X, \lambda)=0,
\end{equation}
whose solution is
\begin{equation}\label{soll}
\Phi(X,\lambda)=e^{iX^{\alpha \beta}\lambda_\alpha \lambda_\beta}\phi(\lambda).
\end{equation}
 At this point, one might ask the question what is the meaning of the equation  (\ref{PRR}) and of the extra coordinates  $y^{mn}$ and $\lambda^\alpha$?
\if{}
 There are two ways to proceed in order to answer this question.

The  first approach is to first integrate out the coordinates $\lambda^\alpha$ and then of the coordinates $y^{mn}$.

 The second approach: This one is reverse to the first one i.e., get rid of the coordinates $y^{mn}$ first and then of the coordinates $\lambda^\alpha$.
\fi
As we shall see,   the answer  is that the equation  (\ref{PRR}) is nothing else but
 Vasiliev's unfolded equations for free massless higher-spin fields in four-dimensional Minkowski space-time \cite{Vasiliev:1988xc}.
The wave function $\Phi(X, \lambda)$ depends on the coordinates $x^m$, $y^{mn}$ and $\lambda^\alpha$. While $x^m$ parameterize the conventional four-dimensional Minkowski space-time, the coordinates $y^{mn}$ (and/or $\lambda^\alpha$) are associated with integer and half-integer spin degrees of freedom of four-dimensional fields with spin values ranging from zero to infinity.

\subsection{Higher spin content of the tensorial space equations} \label{feI}

To demonstrate the above statement let us first Fourier transform the wave function (\ref{soll})
into a conjugate representation with respect to the spinor variable $\lambda_\alpha$ considered in \cite{Vasiliev:2001zy}
\begin{equation}\label{yr}
C(X,\mu)=\int\,d^4\lambda \,e^{-i\mu^\alpha \lambda_\alpha}\Phi(X,\lambda)=\int\,d^4\lambda \,e^{-i\mu^\alpha \lambda_\alpha+
i X^{\alpha \beta}\lambda_\alpha \lambda_\beta}\phi(\lambda).
\end{equation}
The function $C(X,\mu)$ obeys the equation
\begin{equation}\label{Y}
\left( \frac{\partial}{\partial
X^{\alpha \beta}}-
i\frac{\partial^2}{{\partial \mu^\alpha \partial
\mu^\beta}} \right)C(X,\mu)=0.
\end{equation}
Let us expand the function $C(X,\mu)$ in series of the variables $\mu^\alpha$
  \begin{equation}\label{pol}
C(X,\mu)=\sum^\infty_{n=0}C_{\alpha_1\cdots\alpha_n}(X)\,\mu^{\alpha_1}\cdots
\mu^{\alpha_n}=b(X)+f_\alpha (X)\mu^\alpha +\cdots\,.
\end{equation}
and insert this expansion into the equation (\ref{Y}). Then one finds that all the components of  $C(X,\mu)$ proportional to the higher powers
of $\mu^\alpha$ are expressed in terms of  two fields the scalar $b(X)$ and the spinor $f_\alpha(X)$. As a result, of (\ref{pol}) these fields satisfy the relations
\cite{Vasiliev:2001zy}
\begin{eqnarray}\label{b}
\partial_{\alpha \beta}
\partial_{\gamma \delta}\,b(X)-\partial_{\alpha \gamma}\partial_{\beta \delta}\,b(X)&=&0\,,\\
\quad \partial_{\alpha \beta} f_\gamma(X)-\partial_{\alpha \gamma}
f_\beta(X)&=&0\label{f}\,.
\end{eqnarray}
The basic fields $b(X)$ and $f_\alpha(X)$ depend on $x^m$ and $y^{mn}$. Let us now expand these fields
in  series of the tensorial coordinates $y^{mn}$
\begin{eqnarray}
b(x,\,y)&=\phi(x)+y^{m_1n_1}F_{m_1n_1}(x)
+y^{m_1n_1}\,y^{m_2n_2}\,{\hat R}_{m_1n_1,m_2n_2}(x)\nonumber\\
&+\sum_{s=3}^{\infty}\,y^{m_1n_1}\cdots y^{m_sn_s}\,{\hat
R}_{m_1n_1,\cdots,m_sn_s}(x)\,,\label{is}
\end{eqnarray}
\begin{eqnarray}
f^\alpha (x,y)&=&\psi^\alpha(x)+y^{m_1n_1}\,{\hat{\cal R}}^\alpha_{m_1n_1}(x )\nonumber \\
&&+\sum_{s={\frac 52}}^{\infty}\,y^{m_1n_1}\cdots y^{m_{s-{\frac 12}} n_{s-{\frac 12}}}\,{\hat{\cal
R}}^\alpha_{m_1n_1,\cdots,m_{s-{\frac 12}}n_{s-{\frac 12}}}(x )\,.
\label{his}
\end{eqnarray}
Each  four-dimensional component field in this expansion is antisymmetric under the permutation of the indices $m_i$ and $n_i$ and is symmetric
with respect to the permutation of the pairs $(m_i, n_i)$ with $(m_j,n_j)$.
In order to answer the question
about the physical meaning of these fields, let us first consider the scalar field equation (\ref{b}). Using the expression \eqref{tensorial} for the tensorial coordinates and four-dimensional $\gamma$-matrix identities, one can decompose (\ref{b}) as follows
 $$
\partial_p\,\partial^p\,b(x^l,y^{mn})=0, \quad
\left(\partial_p\,\partial_q
-4\,\partial_{pr}\,\partial^r_{~q}\right)\,b(x^l,y^{mn})=0, \quad
\epsilon^{pqrt}\partial_{pq}\,\partial_{rs}\,b(x^l,y^{mn})=0,
$$
\begin{equation} \label{KG}
\epsilon^{pqrt}\partial_q\,\partial_{rt}\,\,b(x^l,y^{mn})=0,
\quad
\partial^{~p}_{q}\,\partial_p\,\,b(x^l,y^{mn})=0\,.
\end{equation}
where $\partial_p = \frac{\partial}{\partial x^p}$ and $\partial_{pq}= \frac{\partial}{\partial y^{pq}}$.
The meaning of the equations (\ref{KG}) is the following.
The first equation is a Klein-Gordon equation. The second equation implies that
the trace (with respect to the $4D$ Minkowski metric) of the tensor which comes with the $s$-th power of $y^{mn}$
in the expansion (\ref{b})
is expressed via the second derivative of the tensor which comes with
the $(s-2)$-th power of $y^{mn}$. Therefore, traces are not
independent degrees of freedom and the independent tensorial fields under consideration are effectively traceless. The third and fourth equation in (\ref{KG})
imply  that the tensor fields   satisfy the four-dimensional Bianchi identities, and the last equation implies  that they are  co--closed.
These are equations for massless higher-spin fields written in terms of their curvatures
${\hat{\cal R}}^\alpha_{m_1n_1,\cdots,m_{s-{\frac 12}}n_{s-{\frac 12}}}(x )$.
In four dimensions these equations are conformally invariant.
Therefore one can conclude that in the expansion \p{is} the field $\phi(x)$ is
 a conformal scalar,  $F_{mn}(x)$ is the field strength of spin-$1$ Maxwell field,
the field ${\hat R}_{m_1n_1,m_2n_2}(x)$ is a linearized Riemann tensor for spin-$2$ graviton, etc.

The treatment of the equation (\ref{f}) which describes half-integer higher-spin fields
in terms of corresponding curvatures is completely analogous to the bosonic one (\ref{b}). The independent
equations for the conformal half-integer spin fields are
\be \label{D4f-1}
\gamma^p \partial_p f (x^l, y^{mn})=0,
\ee
\be \label{D4f-2}
(\partial_p - 2\gamma^r \partial_{pr}) f (x^l,y^{mn})=0
\ee
From \p{D4f-1}--\p{D4f-2} one can derive the equation
\be \label{D4f-3}
\partial_{mn}f (x,y) = \frac{1}{2}\gamma_{[m}\partial_{n]} f(x,y)+\frac{1}{2}(\partial_{mn}+ \frac{1}{2}\varepsilon_{mnpq}\partial^{pq}\gamma_5) f (x,y).
\ee
This equation describes the decomposition of the spinor-tensor $\partial_{mn}f$ into the part which contains the $D=4$ space-time derivative of $f$ and the `physical' part which is
self-dual and gamma-traceless, i.e.
\bea
&&\gamma^m(\partial_{mn}+ \frac{1}{2}\varepsilon_{mnpq}\partial^{pq}\gamma_5) f (x,y)=0 \\ \nonumber
&&(\partial_{mn}+ \frac{1}{2}\varepsilon_{mnpq}\partial^{pq}\gamma_5) f (x^l. y^{mn})=
\frac{1}{2} \varepsilon_{mnrs} (\partial^{rs}+ \frac{1}{2}\varepsilon^{rspq}\partial^{pq}\gamma_5) f (x,y)
\eea
Therefore one can conclude that due to the equations \p{D4f-1}--\p{D4f-2} the field
$\psi^\alpha(x)$ in the expansion \p{his} is a spin-$\frac{1}{2}$ field, the field
${\hat{\cal R}}^\alpha_{m_1n_1}(x )$ corresponds to the field strength of the spin-$\frac{3}{2}$
Rarita--Schwinger field, while the other fields are the field strengths of the half-integer conformal
higher-spin fields in $D=4$.

Finally, let us define the hyperspaces associated with $D=6$ and $D=10$ space--time.
The dynamics of the fields will be again determined by the equation \p{PRE}
with the corresponding hyperspaces and the twistor--like variables $\lambda_\alpha$
defined as follows.

 In $D=10$
 the twistor--like variable $\lambda_\alpha$ is a 16--component
Majorana--Weyl spinor. The gamma--matrices
$\gamma_m^{\alpha \beta}$ and $\gamma_{m_1\cdots m_5}^{\alpha \beta}$ form a basis
of the symmetric $16\times 16$ matrices, so the $n=16$ tensorial
manifold is parameterized by the coordinates
\begin{equation}
X^{\alpha \beta}= {\frac 1{16}}\,\Big(\,x^m\gamma_m^{\alpha \beta}+{\frac 1 {2\cdot
5!}} \,y^{m_1\ldots m_5}\gamma_{m_1\ldots
m_5}^{\alpha \beta}\Big)=X^{\beta \alpha}\,,
\end{equation}
$$
\quad (m=0,1,\ldots,9\,; \quad
\alpha,\beta=1,2,\ldots,16)\,,
$$
where $x^m=X^{\alpha \beta}\gamma^m_{\alpha \beta}\,$ are associated with the
coordinates of the $D=10$ space--time, while the anti--self--dual
coordinates
$$y^{m_1\ldots m_5}=X^{\alpha \beta}\gamma^{m_1\ldots m_5}_{\alpha \beta}=-{\frac 1
{5!}}\,\epsilon^{m_1\ldots m_5n_1\ldots n_5}y_{n_1\ldots
n_5}\,,$$ describe spin degrees of freedom.

The corresponding field equations are again (\ref{b}) and (\ref{f}) and the entire discussion repeats as in the case
of $D=4$. The crucial difference is that now the expansion (\ref{is}) and (\ref{his})
is performed in terms of the coordinates $y^{m_1\ldots m_5}$. As a result
one obtains a description of conformal fields whose curvatures are self--dual
with respect to each set of indexes $(m_i n_i p_i q_i r_i)$.
These traceless rank $5s$ tensors ${ R}_{[5]_1 \cdots [5]_{s} }$  are
automatically irreducible under $GL(10,\mathbb R)$ due to the
self--duality property, and are thus associated with the
rectangular Young diagrams $(s,s,s,s,s)$ which are made of five rows
of equal length $s$ (``multi-five-forms"). The field equations, which are ten--dimensional analogues of the
four-dimensional equations (\ref{KG}), can be found in \cite{Bandos:2005mb}.

In $D =6$ the commuting spinor $\lambda_\alpha$ is  a symplectic Majorana--Weyl
spinor.  The spinor index
can be decomposed as follows
$\alpha=a\otimes i$ ($\alpha=1,\ldots,8$; $a=1,2,3,4$; $i=1,2$). The
tensorial space coordinates $X^{\alpha \beta}=X^{ai\,bj}$ are decomposed
into
\begin{eqnarray}\label{6} && X^{ai\, bj}\, =  \,\frac{1}{8} \, x^m\, {\tilde\gamma}_m^{a
b} \,\epsilon^{ij}\, + \frac{1}{ 16 \cdot 3!}\,y_I^{mnp}\,
{\tilde\gamma}^{ab}_{mnp}\,\tau_I^{ij}\,,\,  \\
& & \qquad m,n,p=0,\ldots,5\,;\quad a,b=1,..., 4\,;\quad
i,j=1,2\,;\quad I=1,2,3\nonumber\end{eqnarray}
where
$\epsilon^{12}=- \epsilon_{12}=1$, and $\tau^{ij}_I$ ($I=1,2,3$)
provide a basis of $2\times 2$ symmetric matrices, They are
related to the usual $SU(2)$-group Pauli matrices $\tau_{I
\, ij}= \epsilon_{jj^\prime}\, \sigma_{I\, i}{}^{j^\prime}$.
  The matrices
${\tilde \gamma}^{ab}_m$ (where $\gamma_{ab}^m= 1/2 \,
\varepsilon_{abcd} {\tilde \gamma}^{m\, cd}$) form a complete
basis of $4\times 4$ antisymmetric matrices with upper (lower)
indices transforming under an (anti)chiral fundamental
representation of the non--compact group $SU^*(4)\sim Spin(1,5)$.
For the space of $4\times 4$ symmetric matrices with upper (lower)
indices,
  a basis is provided by  the set of self--dual
and anti--self--dual matrices   $({\tilde \gamma}^{m n p})^{ab}$
and $\gamma_{ab}^{m n p}$, respectively,
\begin{equation}\label{ggsd}
({\tilde \gamma}^{m n p})^{ab}=\frac{1}{ 3!}\epsilon^{mnpqrs}{\tilde
\gamma}^{ab}_{qrs}\; ,  \qquad \gamma_{ab}^{m n p}=-
\frac{1}{ 3!}\epsilon^{mnpqrs}(\gamma_{qrs})_{ab}\; .
\end{equation}
The coordinates
$ x^m=\,x^{ai\,bj}\,\gamma^m_{ab}\,\epsilon_{ij}$
 are associated with $D=6$ space--time, while the
self-dual coordinates
\begin{equation}\label{6Dy=asd} y_I^{mnp}=
x^{ai\,bj}\,\gamma^{mnp}_{ab}\,\tau_{I\,ij}=- \frac{1}{ 3!}\,\epsilon^{mnpqrs}y^I_{qrs}\; , \end{equation}
describe spinning
degrees of freedom.

The consideration proceeds as in the $D=4$ and $D=10$ case.
Because of the form of the tensorial coordinates in (\ref{6})
 the six-dimensional analogue
of the expansions (\ref{is}) and (\ref{his}) contains powers
of $y_i^{mnp}$. Corresponding field strengths, which again describe conformal fields in six dimensions, are
 self--dual with respect to each set of the indexes $(m_i n_i p_i)$.
 In other words, one has an infinite number of conformally
invariant (self-dual) `multi-3-form' higher-spin fields in the six-dimensional
space--time  which form
the $(2[s]+1)$-dimensional representations of the group $SO(3)$.

In \cite{Gelfond:2003vh,Vasiliev:2007yc,Gelfond:2010pm} the equation \eqref{Y} has been generalized   to include several commuting spinor variables $\mu^{p\alpha}$ $(p,q=1,...,r)$
\begin{equation}\label{Yrr}
\left( \frac{\partial}{{\partial
X^{\alpha \beta}}} \pm i\eta^{pq}\frac{\partial^2}{{\partial \mu^{p \alpha} \partial
\mu^{q \beta}}}\right)C^r_\pm(X,\mu)=0.
\end{equation}
where $\eta^{pq}=\eta^{qp}$ is a nondegenerate metric. The value of $r$ is called the
``rank".
As we explained above the free higher-spin fields in $D=4$ are described by the rank-one
equations in the ten-dimensional tensorial space.
The higher-spin currents are fields of rank-two $r=2$.
These currents obey the equations with off-diagonal $\eta^{pq}$ \cite{Gelfond:2008ur}.
The currents $J(X,\mu^p)$ are bilinear in the higher-spin gauge fields
${\cal C}_+$ and   ${\cal C}_-$,  which obey the rank-one equation
(\ref{Yrr})
$J={\cal C}_+{\cal C}_- $.

On the other hand, when considering rank-two equations the corresponding tensorial
space can be embedded in the higher-dimensional tensorial space.
From the discussion above, it follows that a natural candidate for such higher-dimensional space is  the tensorial extension
of $D=6$ space-time.
In this way one effectively linearizes the problem since the conformal currents in four dimensions are identified with the fields in $D=6$ \cite{Gelfond:2010pm}.

\subsection{Four dimensional unfolded higher-spin field
equations from the hyperspace field equations}
Let us rewrite, in the case of the $D=4$ theory, the hyperspace relations in terms of the Weyl spinors. The momenta \eqref{PRE} take the form
\begin{eqnarray}
P_{AB}=\lambda_A\lambda_B\,,\quad \overline{P}_{\dot{A}\dot{B}}=
\overline{\lambda}_{\dot A}\overline{\lambda}_{\dot B}\,,
\quad P_{A\dot{A}}=\lambda^{}_A\overline{\lambda}_{\dot A}\,,
\end{eqnarray}
while the equation (\ref{PRE}) splits into
\begin{eqnarray}\label{Yy}
\left(\sigma^{mn}_{AB}\frac{\partial}{{\partial
y^{mn}}}+i\frac{\partial^2}{{\partial \mu^A\partial
\mu^B}}\right)C(x,y,\mu)=0,\,\nonumber\\
\\
\left(\overline{\sigma}^{mn}_{{\dot A}{\dot
B}}\frac{\partial}{{\partial y^{mn}}}-i\frac{\partial^2}{{\partial
\overline{\mu}^{\dot A}\partial \overline{\mu}^{\dot
B}}}\right)C(x,y,\mu)=0\nonumber
\end{eqnarray}
and
\begin{eqnarray}\left(\sigma^m_{A\dot{A}}\frac{\partial}{{\partial
x^m}}+i\frac{\partial^2}{{\partial \mu^A\partial \bar{\mu}^{\dot
A}}}\right)C(x,y,\mu)=0\,.\label{unfold}
\end{eqnarray}
Equations (\ref{Yy}) relate the dependence of $C(x,y,\mu)$ on the coordinates
$y^{mn}$ to its dependence on $\mu^\alpha$. Thus using this relation, one can
regard the wave function
$C(x^m,\mu^\alpha):=C(X^{\alpha \beta},\mu^\alpha)|_{y^{mn}=0}$ as the
fundamental field.

The expansion of $C(x^m,\mu)$ in series of $\mu^A$ and
$\overline{\mu}^{\dot A}$
is
\begin{equation}\label{gener}
C(x^p,\mu^A,\overline{\mu}^{\dot
A})=\sum_{m,n=0}^{\infty}\frac{1}{m!n!}\, C_{A_1 \ldots
A_m,\,{\dot B}_1 \ldots {\dot B}_n }(x^p)\, \mu^{A_1} \ldots \mu^{A_m}
\,\overline{\mu}^{{\dot B}_1} \ldots \overline{\mu}^{{\dot B}_n}\,,
\end{equation}
where the reality of the wave function implies $(C_{A_1 \ldots A_m,\,{\dot B}_1 \ldots
{\dot B}_n })^*=C_{B_1 \ldots B_n,\,{\dot A}_1 \ldots {\dot A}_m }$,
and by construction the spin-tensors are symmetric in the indices
$A_i$ and in
$\dot B_i$.

The consistency of (\ref{unfold}) implies the integrability
conditions
\begin{eqnarray}
\label{masseq}\frac{\partial^2}{\partial \mu^{[A}
\partial
x^{B]\dot B}}\, C(x,\mu) =0, \quad
\frac{\partial^2}{\partial \bar{\mu}^{[\dot A} \partial x^{{\dot
B}]B}}\, C(x,\mu) =0\,.
\end{eqnarray}
We have thus obtained the equations of the   Vasiliev's unfolded formulation of free higher spin fields in terms of zero--forms.
In this formulation the
 $C_{0,0}$ component (a physical scalar),
 $C_{A_1 \ldots A_{2s},0}$ and $C_{0, {\dot A}_1, \ldots {\dot A}_{2s}}$ components of the expansion
(\ref{gener}) correspond to the physical fields, while the other fields are auxiliary.
The latter  two fields are
 the self-dual and anti-self-dual
components  of the spin--$s$ field strength. The
nontrivial equations on the dynamical fields are \cite{Vasiliev:1999ba} the
Klein--Gordon equation for the spin zero scalar field  $\partial^m \partial_m
C_{0,0}=0$  and the massless  equations
for spin $s>0$ field strengths
\begin{equation}\partial^{B\dot B}C_{BA_1\ldots A_{2s-1}}(x)=0\,,\quad
\partial^{B\dot B}C_{{\dot B}{\dot A}_1\ldots {\dot A}_{2s-1}}(x)=0\,,\label{Bw}
\end{equation}
which follow from (\ref{masseq}).
All the components of $C(x^m,\mu^A,\overline{\mu}^{\dot
A})$ that depend on {\it both} $\mu^A$ and $\overline{\mu}^{\dot A}$
are auxiliary fields expressed by (\ref{unfold}) in terms of
space--time derivatives of the dynamical fields contained in the
analytic fields $C(x^m,\mu^A,0)$ and $C(x^m,0,\mu^{\dot A})$
and thus one arrives at the unfolded formulation of
\cite{Vasiliev:1999ba}.

Let us summarize what we have considered by now. To describe the dynamics of higher-spin fields in four dimensions
we have introduced extended ten-dimensional tensorial space, hyperspace, parameterized by the coordinates
$X^{\alpha \beta}$ (\ref{tensorial}). The main object is a generating functional for higher-spin fields
described by $C(X, \mu)$ or by $\Phi(X, \lambda)$. The generating functional depends on the tensorial coordinates
$X^{\alpha \beta}$ and on the commuting spinors $\mu^\alpha$ or $\lambda^\alpha$.
The dynamics is described by the field equations (\ref{PRR}) or (\ref{Y}).
To obtain from these the higher-spin field equations in the ordinary space-time parameterized by the coordinates $x^m$
one can use two options. In the first approach one gets rid of the
tensorial coordinates $y^{mn}$ and arrives at Vasiliev's unfolded formulation
in terms of the functional (\ref{gener}). Alternatively, one can first get rid of
the commuting spinor variables and arrive at the  equations
for the bosonic (\ref{is}) and fermionic (\ref{his}) hyperfields.
Both pictures provide the equations for the field strengths
of the  higher-spin potentials, the difference being that these field strengths are realized either as tensors or spin-tensors.

\subsection{Generalized conformal group $Sp(2n)$ }\label{glcg}

Let us consider in more detail the symmetries of the equation (\ref{PRE}) in which now
 the Greek indices $\alpha, \beta,\ldots$ run from 1 to an arbitrary even integer $2n$.
However, as we explained in the previous Section, the physically interesting cases are associated with
$n=2,4, 8,$ and $ 16$, which correspond to the number of space-time dimensions equal to
$3,4,6$ and $10$, respectively.

It turns out that
the equation \p{PRE} is invariant under
the transformations of the $Sp(2n)$ group \cite{Vasiliev:2001dc,Plyushchay:2003gv}
\begin{equation}\label{trr}
\delta\lambda_\alpha=g_\alpha^{~\beta}\lambda_\beta-k_{\alpha \beta}X^{\beta \gamma}\lambda_\gamma,
\end{equation}
\be\label{deltaX}
\delta X^{\mu\nu}= a^{\mu\nu} + (X^{\mu\rho}g_\rho{}^\nu+X^{\nu\rho}g_\rho{}^\mu)-X^{\mu\rho}k_{\rho\lambda}X^{\lambda\nu}\,.
\ee
The constant parameters $a^{\alpha \beta}=a^{ \beta \alpha },$ $g_\gamma^{~\alpha}$ and  $k_{\alpha \beta}=k_{\beta \alpha }$
correspond to the generators of generalized translations
$P_{\alpha\beta}$,  generalized Lorentz transformations and dilatations $G_\beta^{~\alpha}$ (generated by the $GL(n)$ algebra) and
generalized conformal boosts $K_{\alpha \beta}$.
The differential operator representation of these generators have the form
\be\label{Pmunu}
P_{\mu\nu}=-i\frac{\partial}{\partial X^{\mu\nu}}\equiv-i\partial_{\mu\nu},
\ee
\be\label{GL0}
G_{\nu}{}^{\mu}=-2i X^{\mu\rho}\,\partial_{\rho\nu}
\ee
and
\be\label{K}
K^{\mu\nu}=iX^{\mu\rho}X^{\nu\lambda}\partial_{\rho\lambda}
\ee
These symmetries are the hyperspace counterparts of the conventional Poincar\'e translations, Lorentz rotations, dilatations and conformal boosts of  Minkowski space-time. The generalized Lorentz rotations are generated by the traceless operators $L_{\mu}{}^{\nu}=G_{\mu}{}^{\nu}-\frac 1n \delta^{\nu}_\mu\,G_{\lambda}{}^\lambda$, forming the $SL(n)$--algebra, whereas dilatations are generated by the trace of $G_{\mu}{}^{\nu}$.
The generators \eqref{Pmunu}, \eqref{GL0} and \eqref{K} form the $Sp(2n)$ algebra which plays the role of a generalized conformal symmetry in the hyperspace
\bea\label{sp2n}
&[P_{\mu\nu},\,P_{\rho\lambda}]=0,\qquad [K^{\mu\nu},\,K^{\rho\lambda}]=0,\qquad [G_{\nu}{}^{\mu},G_{\lambda}{}^{\rho}]= i(\delta^{\mu}_\lambda\,G_{\nu}{}^{\rho}-\delta^{\rho}_\nu\,G_{\lambda}{}^{\mu})\,,&\nonumber \\
&[P_{\mu \nu},G_\lambda{}^\rho]=-i(\delta_\mu^\rho P_{\nu \lambda} + \delta^\rho_\nu P_{\mu \lambda} ),\qquad
 [K^{\mu \nu},G_\lambda{}^\rho]=i  ( \delta^\mu_\lambda  K^{\nu \rho}   + \delta^\nu_\lambda  K^{\mu \rho}  )\,,&\nonumber \\
& [P_{\mu \nu},K^{\lambda \rho}]=\frac{i}{4} (  \delta^\rho_\mu G_\nu{}^\lambda +  \delta^\rho_\nu G_\mu{}^\lambda
+  \delta^\lambda_\mu G_\nu{}^\rho +  \delta^\lambda_\nu G_\mu{}^\rho)  \,. &
\eea
From the structure of this algebra, one can see that the flat hyperspace $\mathcal M_n$ can be realized as a coset manifold associated with the translations $P=\frac {Sp(2n)}{K\times\!\!\!\!\supset GL(n)}$  where $K\times\!\!\!\!\!\!\supset GL(n)$ is the semi--direct product of the Abelian group generated by the generalized conformal boosts $K_{\mu\nu}$ and the general linear group.

The generators of the translations, Lorentz rotations and conformal boosts
of the conventional conformal group can be obtained from the $Sp(2n)$ generators as projections onto the
$x$-space, for example $p_m= (\gamma_m)^{\mu \nu}P_{\mu\nu}$, etc.

Let us note that the $Sp(2n)$ algebra can be conveniently realized with the use of the twistor-like variables $\lambda_\alpha$ and their conjugate $\mu^\alpha$
\be\label{lm}
[ \mu^\alpha, \lambda_\beta]= \delta^\alpha_\beta.
\ee
In the twistor representation the generators of the $Sp(2n)$
group have the following form
\be
P_{\alpha \beta}=\lambda_\alpha \lambda_\beta , \qquad  G_\alpha^{~\beta}= \lambda_\alpha \mu^\beta,\qquad
K_{\alpha \beta}=\mu_\alpha \mu_\beta.
\ee

The equations \p{b} and \p{f} are invariant under the $Sp(2n)$ transformations
\p{deltaX}, provided that the fields transform as follows
\begin{equation}\label{sp8fb}
\delta b(X)= -(a^{\mu \nu} \partial_{\mu \nu} + \frac{1}{2}g_\mu{}^\mu + 2 g_\nu{}^\mu X^{\nu \rho}
\partial_{\mu \rho} -
 k_{\mu \nu} (\frac{1}{2} X^{\mu \nu}   + X^{\mu \rho} X^{\nu \lambda}\partial_{\rho \lambda})) b(X)\,,
\end{equation}

\begin{eqnarray}\label{sp8ff} \nonumber
\delta f_\rho(X)&=& -(a^{\mu \nu} \partial_{\mu \nu} + \frac{1}{2}g_\mu{}^\mu + 2 g_\nu{}^\mu X^{\nu \lambda}
\partial_{\mu \lambda} -
 k_{\mu \nu} (\frac{1}{2} X^{\mu \nu}   + X^{\mu \tau} X^{\nu \lambda}\partial_{\tau \lambda})) f_\rho(X) +
\\
&&-(g_\rho{}^\nu - k_{\lambda \rho}X^{\lambda \nu}) f_\nu(X)\,.
\end{eqnarray}
Note that these variations contain the term $\frac{1}{2}(g_\mu{}^\mu-k_{\mu \nu} X^{\mu \nu})$,  implying that the fields have the canonical conformal weight $1/2$. A natural generalization of these transformations to fields of a generic conformal weight $\Delta$ is \cite{Vasiliev:2001zy}
\begin{equation}\label{sp8fbD}
\delta b(X)= -(a^{\mu \nu} \partial_{\mu \nu} + \Delta\,(g_\mu{}^\mu-
 k_{\mu \nu}  X^{\mu \nu}) + 2 g_\nu{}^\mu X^{\nu \rho}
\partial_{\mu \rho} -
 k_{\mu \nu}  X^{\mu \rho} X^{\nu \lambda}\partial_{\rho \lambda}) b(X)\,,
\end{equation}

\begin{eqnarray}\label{sp8ffD} \nonumber
\delta f_\rho(X)&=& -(a^{\mu \nu} \partial_{\mu \nu} +\Delta\,(g_\mu{}^\mu-
 k_{\mu \nu}  X^{\mu \nu}) + 2 g_\nu{}^\mu X^{\nu \lambda}
\partial_{\mu \lambda} -
 k_{\mu \nu} X^{\mu \tau} X^{\nu \lambda}\partial_{\tau \lambda}) f_\rho(X)
\\
&&-(g_\rho{}^\nu - k_{\lambda \rho}X^{\lambda \nu}) f_\nu(X)\,.
\end{eqnarray}

\section{Hyperspace extension of AdS spaces} \label{SectionAdS} \setcounter{equation}0

A hyperspace extension of $AdS_D$ spaces is another coset of the $Sp(2n)$ group.
Recall that the usual $AdS_D$ space can be realized as the coset space\footnote{Here, $K$ and
$\mathbb D$ denote the generalized conformal boosts and dilatation, respectively.}
$\frac{SO(2,D)}{K \times\!\!\!\!\supset (SO(1,D-1)\times \mathbb D)}$
parameterized by the  coset element $e^{{\cal P}_m\,x^m}$.
 The generators of the
$AdS_D$ boosts ${\mathcal P}_m$ can be singled out from the generators of the four
dimensional conformal group $SO(2,D)$ by taking a linear
combination of the generators of the Poincar\'e translations $P_m$ and
conformal boosts $K_m$ as ${\cal P}_m=P_m- \xi^2 K_m$, where $\xi$ is the inverse of the $AdS_D$ radius.

Analogously, for the case of the hyperspace  extension of the $AdS_D$ space
let us consider the generators
\begin{equation}\label{6.1}
 {\cal P}_{\alpha \beta}= P_{\alpha \beta}- \frac{\xi^2}{16}K_{\alpha \beta}, \quad [{\cal P},{\cal P}]\sim
M,
 \quad [{\cal P},M]\sim {\cal P},
\end{equation}
 where $K_{\alpha\beta}=C_{\alpha\gamma}C_{\beta\delta}K^{\gamma\delta}$, $M_{\alpha \beta}$ stands for the symmetric part of the $GL(n)$
 transformations $M_{\alpha\beta}=G_{(\alpha}{}^\gamma C_{\gamma \beta)} \equiv \frac 12(G_{\alpha}{}^{\gamma}C_{\gamma\beta}+G_{\beta}{}^{\gamma}C_{\gamma\alpha})$ and $C_{\alpha\beta}=-C_{\beta\alpha}$ is the $Sp(n)$-invariant symplectic metric.
 One can see that the corresponding manifold is an $Sp(n)$ group manifold
 \cite{Plyushchay:2003gv} which can be realized
as a coset space $\frac{Sp(2n)}{K \times\!\!\!\!\supset GL(n)}$ with the coset
element $e^{(P-\frac{\xi^2}{16}K)_{\alpha \beta}\,X^{\alpha \beta}}$.
Indeed, let us recall that $Sp(n)$ group is generated by $n \times n$ symmetric
matrices $M_{\alpha \beta}$ which form the algebra
\begin{equation}\label{algebrasp8}
\left[ M_{\alpha \beta}, M_{\gamma \delta} \right] = - \frac{i\xi}{2} \left[ C_{\gamma (\alpha} M_{\beta) \delta}+
 C_{\delta (\alpha} M_{\beta) \gamma} \right], \quad \alpha, \beta= 1,...,n\,.
\end{equation}

As a group manifold, $Sp(n)$ is the coset  $[Sp(n)_L\times Sp(n)_R]/Sp(n)$ which has the isometry group $Sp(n)_L\times Sp(n)_R$, the latter being the subgroup of $Sp(2n)$ generated by
\be\label{LR}
M^L_{\alpha\beta}= P_{\alpha\beta}-\frac{\xi^2}{16}K_{\alpha\beta}- \frac{\xi}{4}M_{\alpha\beta}\,\qquad M^R_{\alpha\beta}= P_{\alpha\beta} -  \frac{\xi^2}{16}  K_{\alpha\beta} +  \frac{\xi}{4} M_{\alpha\beta}\,,
\ee
as one may see from the structure of the $Sp(2n)$ algebra \eqref{sp2n}.    The generators $M_{\alpha\beta}$ form the diagonal $Sp(n)$ subalgebra of $Sp(n)_L\times Sp(n)_R$.

Let us note that, for the case of $n=4$, i.e., for the case of four space-time dimensions,  $AdS_4$ space is a coset subspace of $Sp(4)\sim SO(2,3)$ of the maximal dimension. For $n>4$, an $AdS_D$ space is also a subspace of $Sp(n)$ manifold but is no longer the maximal coset of this group.

\subsection{GL-flatness of $Sp(n)$ group manifolds} \label{secgl4}

Let us describe a property of $GL$-flatness of the $Sp(n)$ group manifolds which is a generalization of the conformal flatness property of $AdS_D$ spaces.
By $GL$-flatness we mean that, in a local coordinate basis associated with $X^{\alpha\beta}$,
the corresponding $Sp(n)$ Cartan form $\Omega^{\alpha \beta}$ has the form
\begin{equation}\label{sol1}
\Omega^{\alpha \beta} = d X ^{\mu \nu} G_{\mu}{}^\alpha(X) G_{\nu}{}^\beta(X)\,,
\end{equation}
with the matrix $G_{\mu}{}^\alpha(X) $ being
\begin{equation} \label{sol3}
G_\mu{}^{ \alpha}(X) = \delta^\alpha_\mu + \sum_{k=1}^\infty { \left( -\frac{\xi}{4} \right) }^k (X^k)_\mu{}^\alpha\,.
\end{equation}
This expression implies that the $Sp(n)$ Cartan form is obtained from the flat differential $dX^{\mu\nu}$ by a specific $GL(n)$ rotation of the latter.

This property can be  demonstrated  by showing that
the Cartan forms \p{sol1} satisfy the $Sp(n)$-group Maurer-Cartan equations (see \cite{Florakis:2014kfa},
\cite{Plyushchay:2003gv} for technical details)
\be \label{MC11}
d \Omega^{\alpha \beta} +\frac{\xi}{2} \Omega^{\alpha \gamma} \wedge \Omega_\gamma{}^\beta =0\,.
\ee
 The  matrix $G_{\alpha}^{-1 \mu}(X)$ inverse to \eqref{sol3}
 depends linearly on $X_{\alpha}{}^{\mu}$ and has a very simple form
\begin{equation}\label{sol2}
G_\alpha^{-1 \mu}(X) = \delta_\alpha^\mu +\frac{\xi}{4}X_\alpha{}^\mu\,.
\end{equation}
Note that the possibility of representing the Cartan forms in the form \eqref{sol1} is a particular feature of the $Sp(n)$ group manifold since, in general, it is not possible to decompose the components of the Cartan form into a ``direct product" of components of some matrix  $G_{\mu}{}^\alpha$.

\subsection{An explicit form of the $AdS_4$ metric}
Let us now ~demonstrate  ~that, ~for ~the ~case of $n=4$ ($D=4$),  the pure $x^m$-dependent part
of the matrix $G_{\mu}{}^\alpha(X)$ indeed generates the metric on $AdS_4$ in a specific parameterization. 
To this end, we should evaluate the expression
\begin{equation} \label{EX}
\Omega^{\alpha \beta}(x^m) =\frac{1}{2} dx^m {(\gamma_m)}^{\delta \sigma}G_\delta{}^{\alpha} G_{\sigma}{}^\beta =
\frac{1}{2} dx^m e_m^a {(\gamma_a)}^{\alpha \beta} + \frac{1}{4} dx^m \omega_m^{ab} {(\gamma_{ab})}^{\alpha \beta},
\end{equation}
 where the dependence of the matrices $X^{\alpha \beta}$ on the coordinates $y^{mn}$  (see eq. \p{tensorial}) was discarded, i.e. $X_\alpha{}^\beta = \frac{1}{2}x^n (\gamma_n)_\alpha{}^\beta$. Denoting
 \be
 x^2 = x^m x^n \eta_{mn}, \quad  x_m = \eta_{mn} x^n
 \ee
 and, using the explicit form (\ref{sol3}) of $G_\mu{}^\alpha(X)$, one obtains
\begin{align}
	\Omega^{\alpha\beta}(x)= \frac{1}{2} \frac{dx^m}{[1-(\frac{\xi}{8})^2 x^2]^2}\left[(\gamma_\ell)^{\alpha\beta}\left([1+(\tfrac{\xi}{8})^2 x^2]\delta_m^\ell-2(\tfrac{\xi}{8})^2 \eta_{mn}x^n x^\ell\right)-\tfrac{\xi}{4}x^n (\gamma_{mn})^{\alpha\beta}\right] \,.
\end{align}
In this way, we obtain a four-dimensional space vierbein and spin-connection
\begin{align}
	e^a_m &= \frac{1}{[1-(\frac{\xi}{8})^2 x^2]^2}\left( [1+(\tfrac{\xi}{8})^2 x^2]\delta_m^a-2(\tfrac{\xi}{8})^2
x^a x_m\right) \,,\\
	\omega^{ab}_{m} &= \frac{-2\xi}{[1-(\frac{\xi}{8})^2 x^2]^2}\,\delta^{[a}_m x^{b]} = - \frac{8 (\frac{\xi}{8})}{(1-(\frac{\xi}{8})^2 x^2)^2}(x^a \delta^b_m-x^b \delta^a_m)\,.
\end{align}
The corresponding metric is
\be\label{OurMetric}
	g_{mn}= \frac{1}{[1-(\frac{\xi}{8})^2 x^2]^4}\left( [1+(\tfrac{\xi}{8})^2 x^2]^2 \eta_{mn}-4 (\tfrac{\xi}{8})^2 x_m x_n \right) \,,
\ee
It is well-known (see also subsection \ref{scalaronads}) that the metric on $AdS_D$   can be represented as an embedding in a flat $(D+1)$-dimensional
space
\begin{align}\label{embedmetric}
	ds^2 = \eta_{mn} dy^m dy^n - (dy^D)^2~,
\end{align}
via the embedding constraint
\begin{align}
	\eta_{mn}y^m y^n - (y^D)^2 = -r^2~.
\end{align}
Choosing the embedding coordinates for $AdS_4$ to be
\begin{align}\label{Fronsdal2}
	& y^m = \frac{1+(\frac{\xi}{8})^2 x^2}{[1-(\frac{\xi}{8})^2 x^2]^2}\, x^m, \qquad
 y^4 = \sqrt{r^2 + x^2 \frac{1+(\frac{\xi}{8})^2 x^2}{[1-(\frac{\xi}{8})^2 x^2]^2}}~,
\end{align}
one readily recovers the metric  (\ref{OurMetric}), with the parameter $\xi$ being related to the $AdS_4$ radius $r$ as follows
\begin{align}\label{relationxi}
	\xi  = \frac{2}{r}~.
\end{align}
Finally, computing the Riemann tensor
\begin{align}
	{R^{ab}}_{mn}= -32(\tfrac{\xi}{8})^2 \frac{1+(\frac{\xi}{8})^2x^2}{[1-(\frac{\xi}{8})^2x^2]^4}\Bigr( [1+
( \tfrac{\xi}{8})^2x^2 ]\delta^{[a}_{m}\delta^{b]}_n + 4(\tfrac{\xi}{8})^2 x^{[a}\delta^{b]}_{[m}x_{n]} \Bigr)~,
\end{align}
and  the Ricci scalar
\begin{align}
	R = - 192 \left(\frac{\xi}{8}\right)^2 = - 3\xi^2~,
\end{align}
one verifies that the metric   (\ref{OurMetric}) indeed  corresponds to a  space with constant negative curvature, i.e. the $AdS_4$ space.

\section{Particles in hyperspaces} \label{Sectionpareticle} \setcounter{equation}0

In this Section, we would like to explain the physical meaning of the tensorial space coordinates as spin degrees of freedom from the perspective of the dynamics of a particle in hyperspace.

Historically, the first dynamical system in which the Fronsdal
hyperspace proposal for higher--spin fields was realized explicitly was the
twistor--like superparticle model of Bandos and Lukierski
\cite{Bandos:1998vz} which, for $D=4$, possesses the generalized superconformal symmetry under $OSp(1|8)$. The original motivation behind this model
was   a geometric interpretation of commuting tensorial charges in an extended supersymmetry algebra.
Its higher--spin content was found
later in \cite{Bandos:1999qf,Bandos:1999rp} where the quantum
states of the superparticle were shown to form an infinite tower of
massless higher--spin fields, and the relation of this model to the unfolded formulation was assumed. This relation was analyzed in
detail in \cite{Vasiliev:2001zy,Vasiliev:2001dc,Plyushchay:2003gv,Plyushchay:2003tj,Bandos:2005mb}.
In addition to the relation to higher spins,
the model of Bandos and Lukierski \cite{Bandos:1998vz} has revealed other
interesting features, such as the invariance under supersymmetry
with tensorial charges (which are usually associated with brane
solutions of Superstring and M--Theory). Moreover, it has provided the first
example of a dynamical BPS system preserving more than half of the bulk
supersymmetries. BPS states preserving $\frac{2n-1}{2n}$ supersymmetries
(with $n=16$ for $D=10,11$) were then shown to be building
blocks of any BPS states, and this led to a natural conjecture that they can be
elementary constituents or `preons' of M--theory
\cite{Bandos:2001pu}.

Let us consider the generic case of a particle moving in an $Sp(2n)$--invariant hyperspace $\mathcal M$ described by the
 action
\begin{equation}\label{action22}
S[X,\lambda]=\int\,
E^{\alpha\beta}\left(X(\tau)\right)\,\lambda_\alpha(\tau)\,\lambda_\beta(\tau),
\end{equation}
where $X^{\mu\nu}(\tau)$ are the hyperspace coordinates of the particle. The auxiliary
commuting variables
$\lambda_\alpha(\tau)$ $(\alpha=1,\cdots, n)$ is a
real spinor with respect to $Sp(n)$ and a vector with
respect to $GL(n)$ (introduced in Section
\ref{FH}). Finally
 $E^{\alpha\beta}(X(\tau))=E^{\beta\alpha}(X(\tau))=dX^{\lambda\rho}(\tau)E_{\alpha\beta}^{\mu\nu}(X)$ is the pull--back on the particle worldline
of the hyperspace vielbein. For  flat hyperspace
\begin{equation}\label{Omegaf}
E^{\alpha\beta}(X(\tau))=d\tau\,\partial_\tau
X^{\alpha\beta}\,(\tau)=dX^{\alpha\beta}\,(\tau),\,
\end{equation}
and for the case of the $Sp(n)$ group manifold
\be\label{Omega1}
E^{\alpha\beta}(X(\tau))=\Omega^{\alpha\beta}(X),
\ee
where $\Omega^{\alpha \beta}$ is an $Sp(n)$ Cartan form. The latter can be taken
 in the
  $GL$-flat realization as in \eqref{sol1}.
The dynamics of particles on the $OSp(N|n)$ supergroup manifolds
was considered for $N=1$ in \cite{Bandos:1999pq,Plyushchay:2003gv,Plyushchay:2003tj} and for generic values of $N$ in
\cite{Vasiliev:2001zy,Vasiliev:2001dc}, and, as we have already mentioned,
the twistor-like superparticle in the $n=32$ super-hyperspace  was
considered in \cite{Bandos:2003us} as a point-like model for BPS
preons \cite{Bandos:2001pu}, the hypothetical
$\frac{31}{32}$-supersymmetric constituents of M-theory.

The action (\ref{action22}) is manifestly invariant under global
$GL(n)$ transformations and {\it implicitly} invariant under
global $Sp(2n)$ transformations, acting linearly on $\lambda_\rho$ and
non--linearly on $X^{\rho\nu}$. Thus, the model possesses the
symmetry that Fronsdal proposed as an underlying symmetry of
higher--spin field theory in the case $n=4$, $D=4$
\cite{Fronsdal:1985pd}. To make the $Sp(2n)$ invariance
manifest, it is convenient to rewrite the action (\ref{action22}) in a
twistor form (for simplicity we consider the flat case
\p{Omegaf})
\begin{equation}\label{actionmulambda}
S[\lambda,\mu]=\int\,
\left (d\mu^\alpha(\tau)\,\lambda_\alpha(\tau)-\mu^\alpha(\tau)\,d\,\lambda_\alpha(\tau) \right )
=\int\,
dZ^{\mathcal A}Z_{\mathcal A}\,,
\end{equation}
where
\be\label{muxlambda}
\mu^{\alpha}=X^{\alpha\beta}\,\lambda_\beta\,,
\ee
and
\be\label{Z}
Z_{\mathcal A}=(\lambda_\alpha,\,\mu^\beta)\,\qquad Z^{\mathcal A}=C^{\mathcal
{AB}}\,Z_{\mathcal B}=(\mu^\alpha,\,-\lambda_\beta), \quad {\mathcal A}=1,\cdots,
2n\,,
\ee
form a linear representation of $Sp(2n)$
\be\label{ospm}
 \delta Z_{\mathcal A}=S_{\mathcal A}{}^{\mathcal
B}\,Z_{\mathcal B}, \qquad S_{\mathcal A}{}^{\mathcal B}=
 \left( \ba{cc}
g^{~\beta}_{\alpha}  & k_{\alpha\gamma} \\
a^{\delta\beta} & -(g^{~\delta}_{\gamma})^T \\
\ea
\right)\,.
 \ee
Hence, the bilinear form
$dZ^{\mathcal A}\, Z_{\mathcal A}$ is manifestly $Sp(2n)$ invariant. Note that,
as it follows from the action (\ref{actionmulambda}), the variables $\mu^\alpha$
and $\lambda_\beta$ are canonically conjugate coordinates and momenta of the
particle. Upon quantization, they become the operators introduced in Section \ref{glcg}, eq. \eqref{lm}.

Using the relation (\ref{muxlambda})
one can easily recover the $Sp(2n)$ transformation (\ref{deltaX}) of $X^{\alpha\beta}$.

Applying the Hamiltonian analysis to the particle model described by
(\ref{action22}) and (\ref{Omegaf}), one finds that the momentum
conjugate to $X^{\alpha\beta}$ is related to the twistor--like variable
$\lambda_\alpha$ via the constraint
\begin{equation}\label{Penroselike}
P_{\alpha\beta}=\lambda_\alpha \lambda_\beta\,.
\end{equation}
As we have already mentioned, this expression,  \emph{e.g.} in the case $n=4$ for which $X^{\alpha\beta}$ is given in (\ref{tensorial}), is the direct analog and the generalization of the
Cartan-Penrose (twistor) relation for the particle momentum
$P_m=\bar \lambda\,\gamma_m\,\lambda$. A difference is that in $D=4$ the Penrose
twistor relation is invariant under the phase transformation
\be\label{phasetr}
\lambda_\alpha
~\rightarrow~e^{i\varphi\,\gamma^5}\,\lambda_\alpha ,
\ee
or in the two--component
Weyl spinor notation $\lambda_A~\rightarrow~e^{i\varphi}\,\lambda_A$, while eq.(\ref{Penroselike}) does not possess this
symmetry. rather the  symmetry of the
model is
$\mathbb Z_2$ ($\lambda_\alpha~\rightarrow~-\lambda_\alpha$)
subgroup of $U(1)$
and as a result in the
model under consideration the phase component $\varphi$ of $\lambda_\alpha$
is a dynamical degree of freedom. It turns out that upon quantization it is associated with the infinite number of massless quantum  states (particles) with increasing spin (helicity). This is in contrast to the conventional twistor--like (super)particle models with a finite number of quantum states, considered  \emph{e.g.} in \cite{Ferber:1977qx,Shirafuji:1983zd,Bengtsson:1987ap,Bengtsson:1987si,Sorokin:1989zi,Volkov:1988vf,Plyushchay:1989sp,Sorokin:1989jj,Gumenchuk:1990db,Bandos:1990ji,Bandos:1990pk,Plyushchay:1990eq}.

To understand the physical meaning of the phase $\varphi$, let us
notice that eq. (\ref{Penroselike}) is a constraint on  possible
values of the canonical momenta of the particle in the hyperspace.
In the case $n=4$ the Majorana spinor $\lambda_\alpha$ has four
independent components.  One of these components can be associated with the phase
$\varphi$. The momentum $P_m=\bar \lambda\,\gamma_m\,\lambda$ of the particle
along the four conventional Minkowski directions $x^m={1\over
2}\,X^{\mu\nu}\,\gamma_{\mu\nu}^m$ of the hyperspace
(\ref{tensorial}) is light--like. Therefore, $P_m$ depends on three components of
$\lambda_\alpha$. It does not depend on the phase $\varphi$ of $\lambda_\alpha$, since it is invariant under the phase transformation \eqref{phasetr}.
The momentum $P_{mn}=\bar \lambda\,\gamma_{mn}\,\lambda$ of the particle along
the six additional tensorial directions $y^{mn}={1\over
4}\,X^{\alpha\beta}\,\gamma^{mn}_{\alpha\beta}$ is not invariant under the phase transformations and, hence, depends on the four components
of $\lambda_\alpha$. However, we have already associated three of them with the
light--like momentum $P_m$ in $D=4$. Therefore, the only independent component of the momentum
$P_{mn}$ is associated with the $U(1)$ phase $\varphi$ of $\lambda_\alpha$,
and as a result the motion of the particle along the six tensorial
directions $y^{mn}$ is highly constrained. This means that,
effectively, the particle moves in the four-dimensional Minkowski space and
along a single direction in the six additional dimensions whose
coordinate is conjugate to the compact momentum--space direction
parameterized by the periodic phase $\varphi$. As shown in
\cite{Bandos:1999qf,Bandos:1999rp}, the coordinate conjugate to
the compactified momentum $\varphi$ takes, upon quantization, an
infinite set of integer and half--integer values associated with the
helicities of higher--spin fields.  The half--integer and
integer--spin states are distinguished by the discrete symmetry
$\mathbb{Z}_2$ ($\lambda_\alpha~\rightarrow~-\lambda_\alpha$).

The resulting  infinite tower
 of discrete higher--spin states can be regarded \cite{Bandos:1999qf,Bandos:1999rp}
as an alternative to the Kaluza--Klein compactification mechanism akin to
Fronsdal's original proposal. In contrast to the conventional Kaluza--Klein
theory, in the hyperspace particle model, the compactification occurs
in momentum space and not in coordinate space. The
phase $\varphi$ in \p{phasetr}
can be regarded as a compactified component of the
momentum (\ref{Penroselike}), while the corresponding conjugate hyperspace
coordinate is quantized and labels the discrete
values of spin of fields in the effective conventional space--time.

As we have already seen  by virtue of the
Fierz identity (\ref{D101})  the twistor particle momentum is
light--like ($P^m P_m=0$) in $D=3,4,6$ and $10$. Therefore, in the
hyperspaces corresponding to these space--time dimensions the
first--quantized particles are massless \cite{Bandos:1998vz,Bandos:1999qf,Bandos:1999rp}. Moreover,
since the model is invariant under the generalized conformal group
$Sp(2n)$, the quantum states of this particle in the hyperspaces
containing the $D=3,4,6$ and $10$ Minkowski spaces as subspaces correspond
to the conformal higher--spin fields introduced in Section \ref{FH}.

Let us conclude this section with a brief comment on the model describing a particle propagating
on the $Sp(n)$ group manifold. Its action has the form \eqref{action22},
with the corresponding Cartan form given by
\eqref{Omega1}.
The property of $GL$-flatness greatly simplifies the
analysis of this case.
Namely, since the Cartan forms of the $Sp(n)$ group manifold  and the flat hyperspace are
related as in eq. \p{sol1}, one can simply reduce the classical $Sp(n)$ action to the flat one by redefining the spinor variables as follows $\lambda_\alpha \rightarrow G_\alpha^{-1 \beta}(X) \lambda_\beta $. However, when quantizing this system we should work with variables that appropriately describe the geometry of the $Sp(n)$ background in which the particle propagates. Thus upon quantization one gets the eq. \eqref{lsp}
as explained in detail in \cite{Plyushchay:2003tj}.

\section{Field equations on $Sp(n)$ group manifold} \setcounter{equation}0
\label{sectionfesp}
\subsection{Scalar field on $AdS_D$. A reminder} \label{scalaronads}

Before deriving the field equations of hyperfields on $Sp(n)$ group manifolds let us recollect some well known
facts about a scalar field propagating on $AdS_D$ background. In the next subsection we will see that the form of the scalar field equation on  $Sp(n)$ and its certain solutions are somewhat similar to those of the $AdS$ scalar.

Conformally invariant scalar on $AdS_4$
is described by the field equation \cite{Fronsdal:1975ac}
\begin{equation} \label{confsc}
\left ( D^m D_m + \frac{2}{r^2} \right ) \phi(x)=0,
\end{equation}
here $D_m$ is the usual covariant derivative on $AdS_4$.

The equation (\ref{confsc}) can be written in a so-called ambient space formalism. The ambient space is obtained by introducing one more time-like dimension and considering $AdS_D$ as a
 hyperboloid in this higher dimensional space
\footnote{For applications of this formalism to the  description of higher-spin fields
on $AdS_D$ see for example  \cite{Metsaev:1994ys,Metsaev:1995re,Metsaev:1997nj,Metsaev:1998xg,Fotopoulos:2006ci,Bekaert:2012vt,
Bekaert:2013zya}}
\begin{equation}\label{YA}
\eta_{AB}y^A y^B= -r^2, \quad \eta_{AB}= diag(-1,1,..,1,-1), \quad A=0,1,..,D\,.
\end{equation}
The $AdS_D$ ambient-space generalization of \p{confsc} has the form
\begin{equation} \label{adsscalarD}
\left ( \nabla^A \nabla_A + \frac{2(D-3)}{r^2} \right ) \phi(y)=0,
\end{equation}
where
\begin{equation}
\nabla^A = \theta^{AB} \frac{\partial}{\partial y^B}
\end{equation}
and
\begin{equation}
\theta^{AB} = \eta^{AB} + \frac{y^A y^B}{r^2}
\end{equation}
is a projector, since in view of the relation \p{YA}
one has
\begin{equation}
\theta^{AB}\theta^{BC}= \theta^{AC}, \quad y^A \theta_A^B=0, \quad y^A  \nabla_A=0, \quad \nabla^A y_A =D,
\end{equation}
where the indexes $A,B$ are raised and lowered with the metric $\eta^{AB}$ and $\eta_{AB}$.

One also has the following identities
\begin{equation} \label{ambientalg}
[\nabla_A, \nabla_B]= -y_A \nabla_B + y_B \nabla_A, \quad [\nabla^C \nabla_C, y^A]= 2 \nabla^A + D y^A,
\end{equation}
$$
 [\nabla^C \nabla_C, \nabla^A]= (2-D) \nabla^A + 2 y^A \nabla^D \nabla_D\,
$$
where we have set $r^2=1$.
The generators of the $SO(2,D-1)$ group can be expressed as
\begin{equation} \label{GEN}
M^{AB}= y^A \nabla^B - y^B \nabla^A.
\end{equation}
One can  check that
the generators (\ref{GEN}) can also be represented
as
\begin{equation} \label{gen}
M_{AB} = y_A \partial_B - y_B \partial_A, \quad \partial_A = \frac{\partial}{\partial y^A}.
\end{equation}
To form the $SO(2,D)$  conformal algebra we need extra generators.
These generators are
\begin{equation}\label{5A}
M_{(D+1) A} = \partial_A + y_A y^B \partial_B + l y_A
\end{equation}
Here $l$ is the conformal weight of a field. For the scalar $l=1$.

One can derive \p{5A} as follows. Obviously \p{YA} is invariant under the $SO(2,D-1)$ rotations.
In order to realize the conformal transformations in the ambient space one adds to it one more dimension
i.e., considers $D+2$ dimensional space, parameterized by the coordinates $z^M$, where $M=0,1,.., D+1$. These coordinates are subject to the constraint
\begin{equation}\label{constz}
-{(z^0)}^2 + {(z^1)}^2 + {(z^2)}^2 +\cdots+{(z^{D-1})}^2 -  {(z^{D})}^2 + {(z^{D+1})}^2= z^M z^N g_{MN}=0
\end{equation}
which is invariant under the group of rotations $SO(2,D)$ with the generators
\begin{equation} \label{GM}
M_{MN}= z_M \partial_N - z_N \partial_M.
\end{equation}
One can solve the constraint (\ref{constz}) by introducing
\begin{equation}
y^A = r\frac{z^A}{z^{D+1}},
\end{equation}
satisfying eq. \eqref{YA}.

The generators $M_{MN}$ (\ref{GM}) contain the generators $M_{AB}$ of the
$AdS_D$ isometry group $SO(2,D-1)$ and the generators
$M_{(D+1),A}$ which extend the latter to the conformal group $SO(2,D)$ by taking
the functions on the cone (\ref{constz}) to be homogeneous of degree $-l$
\begin{equation}
 z^M \frac{\partial}{ \partial z^M}  f(z) = -l f(z).
\end{equation}
In this way one gets (\ref{5A}).

Then using the explicit realization of the generators (\ref{GEN}), \p{5A}
as well as the commutation relations \p{ambientalg} between the operators   it is  straightforward to check invariance of the field
equation \p{adsscalarD} under the conformal group $SO(2,D)$.

\subsection{$Sp(n)$ group-manifold equations}
In the previous subsection we considered in detail a conformal scalar field on $AdS_D$.
As we discussed in Section \ref{SectionAdS}, the hyperspace generalization of $AdS$ spaces are $Sp(n)$ group manifolds. We will now consider an $Sp(n)$ counterpart of the conformal scalar field equation \eqref{confsc}.

Let us start with an $Sp(n)$ analogue of the equation \p{PRR}.
To this end one should replace the flat  derivative $ \partial_{\alpha \beta}$
with the covariant derivative on $Sp(n)$ group manifold.  The covariant derivatives $\nabla_{\alpha\beta}$
satisfy the $Sp(n)$ algebra
\begin{equation} \label{algebra--11}
[\nabla_{\alpha \beta}, \nabla_{\gamma \delta}] = \frac{\xi}{2}(C_{\alpha (\gamma} \nabla_{\delta)\beta}+ C_{\beta (\gamma} \nabla_{\delta)\alpha }
)\,.
\end{equation}
Due to the $GL$-flatness  these covariant derivatives have a simple form
\begin{equation}\label{nabla}
\nabla_{\alpha \beta} = G_{\alpha}^{-1 \mu}(X) G_{\beta}^{-1 \nu}(X) \partial_{\mu \nu}\,,
\end{equation}
where $G_{\alpha}^{-1 \mu}(X)$ was defined in \eqref{sol2}.
Further, one should  replace the spinor product $\lambda_\alpha \lambda_\beta$ in (\ref{tensorial}) with an expression
which like the covariant derivatives  $\nabla_{\alpha \beta}$
also satisfies the $Sp(n)$ algebra. This can be done by introducing new variables
\be \label{DY}
\tilde Y_\alpha\equiv
\lambda_\alpha+ {i\xi \over 8}{\partial\over {\partial\lambda^\alpha}}
\ee
Obviously, the spinorial variables $Y_\alpha$ do not commute among each other
\be \label{YC}
[\tilde Y_\alpha, \tilde Y_\beta] = \frac{i \xi}{4}C_{\alpha \beta}.
\ee
Using the covariant derivatives $\nabla_{\alpha \beta}$ and the variables
$Y_\alpha$. one can write
an  $Sp(n)$ analogue of the equation (\ref{PRR}) as
\begin{equation}\label{lsp}
\left[\nabla_{\alpha \beta}-{i\over 2}(\tilde Y_\alpha
\tilde Y_\beta +\tilde Y_\beta Y_\alpha)\right]\Phi(X,\lambda)=0\,.
\end{equation}
Similarly, one finds an $Sp(n)$ version of the equation (\ref{Y})
\begin{equation}\label{ysp}
\left[\nabla_{\alpha \beta}-{i\over 2}( Y_\alpha
Y_\beta+ Y_\beta Y_\alpha)\right]C(X,\mu)=0,\,\quad Y_\alpha\equiv
{\frac{\xi} 8}\mu_\alpha+i\frac{\partial}{\partial \mu^\alpha}\,.
\end{equation}
In order to obtain the equations for component fields one should expand, e.g. the functional
$C(X,\mu)$ in power of $\mu^\alpha$
 \begin{equation}\label{pol-1}
C(X,\mu)=\sum^\infty_{n=0}C_{\alpha_1\cdots\alpha_n}(X)\,\mu^{\alpha_1}\cdots
\mu^{\alpha_n}=B(X)+F_\alpha (X)\mu^\alpha +\cdots\,.
\end{equation}
 Plugging this expansion into \eqref{lsp} one can show that similarly to  the case of the flat
 hyperspace only zeroth and the first components in the expansion in terms of the variables $\mu^\alpha$ are independent
 fields whereas the other fields are expressed in terms of derivatives of the independent ones. The independent
 hyperfields $B(X)$ and $F_\alpha (X)$ satisfy
  the equations
  \cite{Plyushchay:2003tj}
\begin{eqnarray}\label{bsp}
\nabla_{\alpha [\beta}\nabla_{\gamma]\delta}B(X)&=&{\xi \over
16}\left(C_{\alpha [\beta }\nabla_{\gamma]\delta}- C_{\delta [\gamma}\nabla_{\beta] \alpha} +
2C_{\beta \gamma}\nabla_{\alpha \delta}\right)B(X)+ \\ \nonumber
&&+{\xi^2\over
64}\left(2C_{\alpha \delta}C_{\beta \gamma}-C_{\alpha [\beta}C_{\gamma]\delta}\right)B(X),
\end{eqnarray}
\begin{equation}\label{fsp}
\nabla_{\alpha[\beta }F_{\gamma]}(X)=-{\xi \over
4}\left(C_{\alpha [\gamma}F_{\beta]}(X)+2C_{\beta \gamma}F_\alpha(X)\right).
\end{equation}
The derivation of these equations
which are  $Sp(n)$ versions of the equations \p{b} and \p{f}
is straightforward  and is given in the Appendix \ref{derivationofeqs}.

Note that if one introduce the covariant derivatives $D_{\alpha\beta}$ acting on the spinors as follows (see \cite{Florakis:2014kfa} for more details)
\be\label{DF}
D_{\alpha\beta}F_\gamma(X)=\nabla_{\alpha\beta}F_\gamma(X)+\frac \xi 4C_{\gamma(\alpha}F_{\beta)}(X)
\ee
the form of the equations \eqref{bsp} and \eqref{fsp} simplifies to
\be\label{bspD}
D_{\alpha [\beta}D_{\gamma]\delta}B(X)={\xi^2\over
8^2}\left(2C_{\alpha \delta}C_{\beta \gamma}-C_{\alpha [\beta}C_{\gamma]\delta}\right)B(X),
\ee
\be\label{fspD}
D_{\alpha[\beta}F_{\gamma]}(X)=0\,.
\ee
We see that eq. \eqref{bspD} reminds that of the $AdS$ scalar field \eqref{confsc}, especially when we contract its indices.

\subsubsection{Connection between the fields in flat hyperspaces  and $Sp(n)$ group manifolds}
One can check \cite{Florakis:2014kfa} using the equations
\be\label{100}
\partial_{\mu\nu}G^{-1\alpha\beta}(X)=\frac\xi 8(\delta_\mu^\alpha\delta^\beta_\nu+\delta_\mu^\beta\delta^\alpha_\nu)\,,
\ee
and
\begin{equation}\label{pG-D}
\partial_{\mu \nu} ({\det G (X)})^{k}= \frac{\xi k}{8} ({\det G(X)})^{k} (G_{\mu \nu}(X) + G_{\nu \mu}(X))\,,
\end{equation}
that the fields  $B(X)$ and $F_\alpha(X)$ satisfying equations \p{bsp}--\p{fsp} are related to the fields $b(X)$ and $f_\mu(X)$ satisfying the flat hyperspace equations \p{b}--\p{f} as follows
\be\label{defB}
B(X)=(\det G(X))^{-\frac 12}\,b(X)\,,
\ee
\be\label{fF}
 F_\alpha(X) =(\det G(X))^{-\frac 12}\, G^{-1}_\alpha{}^\mu(X) f_\mu(X).
\ee
These relations are similar to the relations between the conformally invariant scalar and spinor equations in the conventional flat and $AdS$ spaces and reduce to them in the case of $n=2$, $D=3$.

\subsubsection{Plane wave solutions}
The equations (\ref{lsp})-(\ref{ysp}) can be solved to obtain ``plane-wave" solutions. Let us consider the case
of the $Sp(4)$ group manifold. One can check that
the equations (\ref{lsp})--(\ref{ysp}) have the following  solutions
\begin{equation}\label{ls}
\Phi(X,\lambda)  =\int\, d^4 \mu \, \sqrt{\det G^{-1}(X)}\,
e^{{i}X^{\alpha \beta}(\lambda_\alpha+ {\xi \over 8}\mu_\alpha
)(\lambda_\beta+{\xi \over 8}\mu_\beta) +i \lambda_\alpha \mu^\alpha}\,\varphi(\mu)\,,
\end{equation}
\begin{equation}\label{ys}
C(X,\mu)=\int\, d^4\lambda \, \sqrt{\det G^{-1}(X)}\,
e^{{i}X^{\alpha \beta}(\lambda_\alpha+ {\xi \over 8}\mu_\alpha
)(\lambda_\beta+{\xi \over 8}\mu_\beta) -i \lambda_\alpha \mu^\alpha}\,\varphi(\lambda)\,.
\end{equation}
These solutions describe plane-wave-like fields in the $GL$--flat parameterization of the metric \cite{Plyushchay:2003tj}.
They can be compared with the plane-wave solutions for the higher-spin curvatures
on $AdS_4$ given in \cite{Bolotin:1999fa,Plyushchay:2003gv}.
The latter can be found by solving
the  $AdS_4$ deformation of the field equations \p{Bw}
\bea \label{ads4-1}
&&D_{M \dot M}C_{A_1, ..., A_{n+2s}, \dot A ,..., \dot A_n}(x)= \\ \nonumber
&&= e^{A \dot A}_{M \dot M}C_{A_1, ..., A_{n+2s},A, \dot A ,..., \dot A_n \dot A}(x) -
n(n+2s) e_{M \dot M, \{ A \dot A}C_{A_2, ..., A_{n+2s}, \dot A ,..., \dot A_n }(x)
\eea
where $D_{M \dot M}$ is a covariant derivative on $AdS_4$ and
$e^{A \dot A}_{M \dot M}$ are the corresponding vierbeins in the Weyl spinor representation.
The physical higher-spin curvatures satisfy the equations
\be \label{ads4-2}
e_{A \dot A}^{M \dot M} D_{M \dot M}
C^{A_1, ..., A_{2s}}(x)=0
\ee
whereas the auxiliary fields are expressed via derivatives of the physical fields
with the help of the equation \p{ads4-1}.
Choosing the $AdS_4$ metric in the conformally flat form
\be
e^{A \dot A}_{M \dot M}= e^{\frac{\rho(x)}{2}}\delta^A_M \delta^{\dot A}_{\dot M}, \quad \rho(x)=
\ln{\frac{4}{{(1-{(\frac{x}{r})}^2)}^2}}
\ee
one can find the plane wave solutions of the equation \p{ads4-2}
\be\label{adpws-1}
C_{A_1, ..., A_{2s}}(x)=  \frac{\partial}{\partial \mu^{A_1}}... \frac{\partial}{\partial \mu^{A_{2s}}} C(x, \mu, \overline \mu)|_{\mu=\overline \mu=0}
\ee
with
\bea\label{adpws-2}
&& C(x, \mu, {\overline \mu}) = \int d^2 \lambda d^2 \overline \lambda \Phi(\lambda, \overline \lambda)
\cdot
\\ \nonumber
&& \cdot
\exp{\left( i (\mu_A \overline \mu_{\dot A}+ \lambda_A \overline \lambda_{\dot A}) x^{A \dot A} - \frac{\rho(x)}{2}+
{\left (1-{\left ( \frac{x}{r} \right )}^2 \right)}^\frac{1}{2} (\mu^A \lambda_A + \overline \mu^{\dot A} \overline \lambda_{\dot A} ) \right )}.
\eea
 Comparing \p{adpws-2} with \p{ys}, one can see that the latter is
 a direct generalization of the $AdS_4$ plane-wave solution to the case of the $Sp(4)$ group manifold.

As a simplest example of this construction
let us note that the conformal scalar on $AdS_4$ discussed in  Subsection \ref{scalaronads}
admits a plane-wave solution \cite{Plyushchay:2003gv} of the form
\be \label{scpwads}
\phi(x)= \int d^2 \lambda d^2 \overline \lambda e^{\ i x^{A \dot A}\lambda_A
\overline \lambda_{\dot A} - \frac{1}{2}\rho(x)} \phi_0 (\lambda, \overline \lambda)
\ee
which can be checked substituting the expression \p{scpwads} into the field equation
\p{confsc}.

\subsection{$Sp(2n)$ transformations of the fields}

Using the relation between the fields of weight $\Delta= \frac{1}{2}$ on flat hyperspace and on $Sp(n)$ group manifold  \eqref{defB}
we have the following relation between the $Sp(2n)$ transformations of the weight-$\frac{1}{2}$  fields on $Sp(n)$ and in flat hyperspace
\if{}
\be\label{DB}
B'(X')=({\det G(X')})^{-\frac 12}\,b'(x')\,\quad F'_\alpha(X')=({\det G(X')})^{-\frac 12} \,G_\alpha^{-1\mu}(X')\, f'_{\mu}(X')
\ee
\fi
\be\label{db}
\delta B(X)=({\det G(X)})^{-\frac 12}\delta b(X)\,,
\ee
\be\label{dfF}
\delta F_\alpha(X)=({\det G(X)})^{-\frac 12} \,G_\alpha^{-1\mu}(X)\,\delta f_{\mu}(X)\,.
\ee
Note that in the above expressions the matrix $G_\alpha{}^\mu(X)$ is not varied since it is form-invariant, i.e. $G(X')$ has the same form as $G(X)$.

Then, the $Sp(n)$-variations of $B(X)$ and $F_\alpha(X)$ have the following form \cite{Florakis:2014kfa}
\bea\label{sp8fB-01} \nonumber
&&\delta {B}(X)= -(a^{\alpha \beta} {\cal D}_{\alpha \beta} +\frac 12  (g_\alpha{}^\alpha- k_{\alpha \beta}X^{\alpha \beta})+ 2 g_\beta{}^\alpha X^{\beta \gamma}
{\cal D}_{\alpha \gamma} - \\
 &&-k_{\alpha \beta}  X^{\alpha \gamma} X^{\beta \delta}{\cal D}_{\gamma \delta}) {B}(X)\,,
\eea
\begin{eqnarray}\label{sp8fF-01} \nonumber
&&\delta {F_\sigma}(X)= -(a^{\alpha \beta} {\cal D}_{\alpha \beta} + \frac 12  (g_\alpha{}^\alpha-k_{\alpha \beta} X^{\alpha \beta}) + 2 g_\beta{}^\alpha X^{\beta \gamma}
{\cal D}_{\alpha \gamma} - \\ \nonumber
 &&-k_{\alpha \beta}  X^{\alpha \gamma} X^{\beta \delta}{\cal D}_{\gamma \delta}) {F_\sigma}(X)-(g_\sigma{}^\beta - k_{\sigma \alpha}X^{\alpha \beta}) F_\beta(X),
\end{eqnarray}
where the derivative ${\cal D}_{\alpha \beta}$ is defined as
\begin{equation}\label{cD0}
{\cal D}_{\alpha \beta}= \partial_{\alpha \beta} + \frac{\xi}{16} (G_{\alpha \beta}(X) + G_{\beta \alpha}(X))\,.
\end{equation}
Using
\begin{equation}\label{101}
\partial_{\mu \nu} G_\rho{}^\sigma (X) = \frac{\xi}{8}(G_{\rho \mu}(X)G_\nu{}^\sigma(X) + G_{\rho \nu}(X) G_\mu{}^\sigma (X) )\,,
\end{equation}
one can check that these derivatives commute with each other
$[ {\cal D}_{\alpha \beta}, {\cal D}_{\gamma \delta} ]=0$ just as in the flat case.

Let us note  that the relation between the flat and $Sp(n)$ hyperfields of an arbitrary weight $\Delta$ and the form of the corresponding $Sp(2n)$ transformations require additional study since for this one should know the form of $Sp(2n)$--invariant equations satisfied by these fields, which is still an open problem.

\section{Supersymmetry} \label{SECTIONSUSY} \setcounter{equation}0

In this Section, we present a supersymmetric generalization of the $Sp(2n)$ invariant systems. We will mainly follow \cite{Florakis:2014aaa}.

\subsection{Flat hyper-superspace and its symmetries}\label{fss}

The concept of hyperspaces, hyperfields and of the corresponding field equations can be generalized to construct
supersymmetric $OSp(1|2n)$ invariant systems and the corresponding infinite-dimensional higher-spin supermultiplets. In this section we shall describe this generalization in detail.

The flat hyper--superspace (see \emph{e.g.} \cite{Bandos:1999qf,Vasiliev:2001zy,Bandos:2004nn}) is parameterized by $\frac{n(n+1)}{2}$ bosonic matrix coordinates $X^{\mu\nu}=X^{\nu\mu}$ and $n$ real Grassmann--odd `spinor' coordinates $\theta^\mu$ ($\mu=1,\cdots,n$).
The supersymmetry variation
\begin{equation}\label{susyv}
\delta\theta^\mu=\epsilon^\mu,\qquad \delta X^{\mu\nu}=-i\epsilon^{(\mu}\theta^{\nu)}
\,,
\end{equation}
leaves invariant the Volkov-Akulov-type one-form
\be\label{VA}
\Pi^{\mu\nu}=dX^{\mu\nu}+i\theta^{(\mu}d\theta^{\nu)}\,.
\ee
The supersymmetry transformations form a generalized super--translation algebra
\be\label{ST}
\{Q_\mu,Q_{\nu}\}=2P_{\mu\nu},\qquad [Q_\mu,P_{\nu\rho}]=0\,, \qquad [P_{\mu\nu},P_{\rho\lambda}]=0\,,
\ee
with $P_{\mu\nu}$ generating translations along $X^{\mu\nu}$.

The realization of $P_{\mu\nu}$ and $Q_\mu$ as differential operators is given by
\be\label{PQ}
P_{\mu\nu}=-i\frac{\partial}{\partial X^{\mu\nu}}\equiv-i\partial_{\mu\nu}\,,\qquad Q_{\mu}=\partial_\mu-i\theta^\nu\partial_{\nu\mu}\,,\qquad \partial_\mu\equiv \frac\partial{\partial\theta^\mu}\,,
\ee
The algebra \eqref{ST} is invariant under  rigid $GL(n)$ transformations
\be\label{GL}
Q'_{\mu}=g_{\mu}{}^{\nu}\,Q_\nu\,,\qquad P'_{\mu\nu}=g_{\mu}{}^\rho\,g_\nu{}^\lambda\,P_{\rho\lambda},
\ee
generated by
\be\label{L}
G_{\mu}{}^{\nu}=-{2i}(X^{\nu\rho}+\frac i2 \theta^\nu\theta^\rho)\partial_{\rho\mu}-i\theta^\nu\,Q_\mu\,,
\ee
which act on $P_{\mu\nu}$ and $Q_\mu$ as follows
\be\label{LPLQ}
[P_{\mu \nu},G_\lambda{}^\rho]=-i(\delta_\mu^\rho P_{\nu \lambda}+ \delta^\rho_\nu P_{\mu \lambda} )\,,\qquad [Q_{\mu},G_\nu{}^{\rho}]=-i\delta_{\mu}^{\rho}\,Q_\nu \,,
\ee
and close into the $gl(n)$ algebra as in \eqref{sp2n}
\be\label{gl-1}
[G_{\nu}{}^{\mu},G_{\lambda}{}^{\rho}]= i(\delta^{\mu}_\lambda\,G_{\nu}{}^{\rho}-\delta^{\rho}_\nu\,G_{\lambda}{}^{\mu})\,.
\ee
The algebra \eqref{ST}, \eqref{LPLQ} and \eqref{gl-1} is the hyperspace counterpart of the conventional super--Poincar\'e algebra enlarged by dilatations. That this is so can be most easily seen by taking \hbox{$n=2$} (i.e. $\mu=1,2$), in which case this algebra is recognized as the \hbox{$D=3$} super--Poincar\'e algebra with \hbox{$G_\mu{}^\nu-\frac 12 \delta_\mu^{\nu}\,G_\rho{}^\rho=M_m(\gamma^m)_\mu{}^{\nu}$} ($m=0,1,2$) generating the $SL(2,R)\sim SO(1,2)$ Lorentz rotations
and ${\mathbb D}=\frac 12 G_{\rho}{}^\rho$ being the dilatation generator. Note that the factor $\frac 12$ in the definition of the dilatation generator is required in order to have the canonical scaling of the momentum generator $P_{\mu\nu}$ with weight 1 and the supercharge $Q_\mu$ with weight $\frac 12$, as  follows from eq. \eqref{LPLQ}.

This algebra may be further extended to the $OSp(1|2n)$ algebra, generating generalized superconformal transformations of the flat hyper--superspace, by adding the additional set of supersymmetry generators
\be\label{S-1}
S^\mu=-(X^{\mu\nu}+\frac i2\theta^\mu\theta^\nu)Q_\nu\,,
\ee
and the generalized conformal boosts
\be\label{K-1}
K^{\mu\nu}=i(X^{\mu\rho}+\frac i2\theta^\mu\theta^\rho)(X^{\nu\lambda}+\frac i2\theta^\nu\theta^\lambda)\partial_{\rho\lambda}-i\theta^{(\mu}S^{\nu)}\,.
\ee
The generators $S^\mu$ and $K^{\mu\nu}$ form a superalgebra similar to \eqref{ST}
\be\label{STd}
\{S^\mu,S^{\nu}\}=-2K^{\mu\nu},\qquad [S^\mu,K^{\nu\rho}]=0\,, \qquad [K^{\mu\nu},K^{\rho\lambda}]=0\,,
\ee
while the non--zero (anti)commutators of $S^\mu$ and $K^{\mu\nu}$ with $Q_\mu$, $P_{\mu\nu}$ and $G_\mu{}^\nu$ read
\bea\label{QSP}
&&\{Q_\mu,S^\nu\}=-G_\mu{}^\nu\,,\quad [S^\mu,P_{\nu\rho}]=i\delta^\mu_{(\nu}\,Q_{\rho)}, \\ \nonumber &&[Q_\mu,K^{\nu\rho}]=-i\delta_\mu^{(\nu}\,S^{\rho)}\,,\quad [S^\mu,G_\nu{}^\rho]=i\delta^\mu_\nu\,S^\rho\,.
\eea
Let us note that in the case $n=4$, in which the physical space--time is four--dimensional  the generalized superconformal group $OSp(1|8)$ contains the $D=4$ conformal symmetry group $SO(2,4)\sim SU(2,2)$ as a subgroup, but  not the superconformal group $SU(2,2|1)$. The reason being that, although $OSp(1|8)$ and $SU(2,2|1)$ contain the same number of  (eight) generators, the anticommutators of the former close on the generators of the whole $Sp(8)$, while those of the latter only close on an $U(2,2)$ subgroup of $Sp(8)$, and the same supersymmetry generators cannot satisfy the different anti--commutation relations simultaneously. In fact, the minimal $OSp$--supergroup containing $SU(2,2|1)$ as a subgroup is $OSp(2|8)$.

\subsection{Scalar superfields and their $OSp(1|2n)$--invariant equations of motion}

Let us now consider a superfield $\Phi(X,\theta)$ transforming as a scalar under the super--translations \eqref{PQ}
\be\label{dPhi}
\delta \Phi (X,\theta)=-(\epsilon^\alpha Q_\alpha\,+ia^{\mu\nu}P_{\mu\nu})\,\Phi (X,\theta)\,.
\ee
To construct equations of motion for $\Phi(X,\theta)$ which are invariant under \eqref{dPhi} and comprise the equations of motion of an infinite tower of integer and half-integer higher-spin fields with respect to conventional space--time, we introduce the spinorial covariant derivatives
\be\label{D}
D_\mu=\partial_\mu+i\theta^\nu\partial_{\nu\mu}\,, \qquad \{D_\mu,D_\nu\}=2i\partial_{\mu\nu}\,,
\ee
which (anti)commute with $Q_\mu$ and $P_{\mu\nu}$.

The $\Phi$--superfield equations then take  the  form \cite{Bandos:2004nn}
\be\label{DDPhi}
D_{[\mu}D_{\nu]}\Phi(X,\theta)=0\,,
\ee
As was shown in \cite{Bandos:2004nn}, these superfield equations imply that all the components of $\Phi(X,\theta)$ except for the first and the second one in the $\theta^\mu$--expansion of  $\Phi(X,\theta)$  should vanish
\be\label{bftheta}
\Phi(X,\theta)=b(X)+i\theta^\mu\,f_\mu(X)+i\theta^{\mu} \theta^\nu A_{\mu\nu}(X)+\cdots\,,
\ee
(i.e. $A_{\mu_1\ldots \nu_k}=0$ for $k>1$) while the scalar and spinor fields $b(X)$ and $f_\mu(X)$ satisfy the equations
\p{b}--\p{f}.

The superfield equations \eqref{DDPhi} are invariant under the generalized superconformal $OSp(1|2n)$ symmetry, provided that $\Phi(X,\theta)$ transforms
as a scalar superfield with the `canonical' generalized scaling weight $\frac 12$, \emph{i.e.}
\bea\label{deltaPhi}
\delta\Phi (X,\theta) &=&-(\epsilon^\mu\,Q_\mu+\xi_\mu\, S^\mu+ia^{\mu\nu}\,P_{\mu\nu}+ik_{\mu\nu}\,K^{\mu\nu}+ig_\mu{}^\nu\,G_{\nu}{}^\mu)\,\Phi (X,\theta)\nonumber\\
&&-\frac 12\, \left(g_\mu{}^\mu-
 k_{\mu \nu} (X^{\mu \nu}+\frac i2 \theta^\mu\theta^\nu)+\xi_\mu\,\theta^\mu\right)\,\Phi (X,\theta)\,,
\eea
where the factor $\frac 12$ in the second line is the generalized conformal weight and $\epsilon^\mu$, $\xi_\mu$, $a^{\mu\nu}$, $k_{\mu\nu}$ and $g_\mu{}^\nu$ are the rigid parameters of the $OSp(1|2n)$ transformations.

Scalar superfields with  anomalous generalized conformal dimension $\Delta$ transform under $OSp(1|2n)$  as
\bea\label{deltaPhiD}
\delta\Phi(X,\theta) &=&-(\epsilon^\mu\,Q_\mu+\xi_\mu\, S^\mu+ia^{\mu\nu}\,P_{\mu\nu}+ik_{\mu\nu}\,K^{\mu\nu}+ig_\mu{}^\nu\,G_{\nu}{}^\mu)\,\Phi (X,\theta)\nonumber\\
&&-\Delta\, \left(g_\mu{}^\mu-
 k_{\mu \nu} (X^{\mu \nu}+\frac i2 \theta^\mu\theta^\nu)+\xi_\mu\,\theta^\mu\right)\,\Phi (X,\theta)\,.
\eea
It is instructive to demonstrate how the generalized conformal dimension $\Delta$, which is defined to be the same for all values of $n$ in $OSp(1|2n)$, is related to the conventional conformal weight of scalar superfields in various space--time dimensions. As we have already mentioned in Section \ref{fss}, the dilatation operator should be identified with $\mathbb D=\frac 12 G_{\mu}{}^\mu$. Therefore, considering a $GL(n)$ transformation \eqref{deltaPhiD} with the parameter $g_{\mu}{}^{\nu}$
$$
\delta \Phi (X,\theta) =-ig_\mu{}^\nu\,G_{\nu}{}^\mu\Phi (X,\theta),
$$
the part of the transformation corresponding to the dilatation reads
\be\label{Dt}
\delta_{\mathbb D} \Phi (X,\theta)=-\frac in g_\mu{}^{\mu}\,G_{\nu}{}^\nu\Phi(X,\theta)=-
\frac {2i}n g_{\mu}{}^\mu {\mathbb D}\Phi (X,\theta)=-i \tilde g {\mathbb D}\Phi (X,\theta)\,,
\ee
where $\tilde g=\frac 2n g_{\mu}{}^\mu$ is the genuine dilatation parameter. From \eqref{deltaPhiD}, it then follows that the conventional conformal weight $\Delta_D$ of the scalar superfield is related to the generalized one $\Delta$ via
\be\label{DD}
\Delta_{{ D}}=\frac n2 \Delta\,,  \quad {D}= \frac{n}{2}+2.
\ee
In the $n=2$ case  corresponding to the ${\mathcal N}=1$, $D=3$ scalar superfield theory the two definitions of the conformal dimension coincide, whereas in the case $n=4$ describing conformal higher-spin fields in $D=4$ one finds $\Delta_4=2\Delta$.
 Relation \eqref{DD} indeed provides the correct conformal dimensions of scalar superfields (and consequently of their components) in the corresponding space-time dimensions. For instance, when $\Delta=\frac 12$, in $D=3$ one finds $\frac 12$ as the canonical conformal dimension of the scalar superfield, while in the cases $D=4$  and  $D=6$ ($n=8$) it is found to be equal to one and two, respectively. For convenience, we shall henceforth associate the scaling properties of the fields to the universal $D$-- and $n$--independent generalized conformal weight $\Delta$.

\subsection{Infinite-dimensional higher-spin representation of ${\mathcal N}=1$, $D=4$ supersymmetry}
Using the example of $n=4$ ($D=4$) we will now show that in four space--time dimensions, the fields of integer and half--integer spin $s=0, \frac 12, 1, \cdots, \infty$ encoded in $b(X)$ and $f_\mu(X)$ (see subsection  \ref{feI})
{form an irreducible infinite--dimensional supermultiplet with respect to the supersymmetry transformations generated by the \emph{ generalized} super--Poincar\'e algebra \eqref{ST}.} The hyperfields $b(X)$ and $f_\mu(X)$, satisfying \p{b}--\p{f},
transform under the supertranslations \eqref{dPhi} as follows
\be\label{susybf}
\delta b(X)=-i\epsilon^\mu\,f_\mu(X)\,,\qquad \delta f_\mu(X)= -\epsilon^\nu\,\partial_{\nu\mu}\,b(X)\,.
\ee
and their expansion in terms of the $y^{mn}$ coordinates  is given in \p{is}--\p{his}.

The fact that the higher-- spin fields should form an infinite--dimensional representation {of the generalized} $ \mathcal N=1$, $D=4$ supersymmetry \eqref{ST} is prompted by the observation that the spectrum of bosonic fields contains a single real scalar field $\phi(x)$, which alone cannot have a fermionic superpartner, while each field with $s>0$  has two helicities $\pm s$. Indeed, from \eqref{susybf}, we obtain an infinite entangled chain of supersymmetry transformations for the $D=4$ fields
\bea\label{susycomp} \nonumber
&&\delta\phi(x)=-i\epsilon^\mu\,\psi_\mu (x), \\ \nonumber
&&\delta\psi_\mu(x)= \epsilon^\nu(\gamma^m_{\nu\mu}\,\partial_m\phi(x)
+\gamma^{mn}_{\nu\mu }\,F_{mn}(x)), \nonumber \\
&& \delta F_{mn}(x)=-i\epsilon^\mu \left({ R}_{\mu\,mn}(x)-{1\over
2}\partial_{[m}(\gamma_{n]}\psi)_\mu(x)\right)\,, \\ \nonumber
&&\delta{ R}_{\mu\,mn}(x)= {1\over
2}\partial_{[m}(\gamma_{n]}\delta \psi(x))_\mu-\frac 12\epsilon^\nu\,\gamma^p_{\nu\mu}\,\partial_pF_{mn}(x) - \\ \nonumber &&-\epsilon^\nu\,\gamma^{pq}_{\nu\mu}\left(R_{pq,mn}(x)-{1\over 2}\partial_{q}\eta_{p[m}\partial_{n]}\phi(x)\right),
\eea
and so on.

The algebraic reason behind the appearance of the infinite--dimensional supermultiplet of the $D=4$ higher--spin fields is related to the following fact. In the $n=4$, $D=4$ case the superalgebra \eqref{ST} takes the following form
\be\label{ST4}
\{Q_\mu,Q_\nu\}=(\gamma^m)_{\mu\nu}P_m+(\gamma^{mn})_{\mu\nu} Z_{mn}\,,
\ee
where $P_m$ is the momentum along the four--dimensional space--time and $Z_{mn}=-Z_{nm}$ are the tensorial charges associated with the momenta along the extra coordinates $y^{mn}$.

On the other hand, the conventional $N=1$, $D=4$ super--Poincar\'e algebra is
\be\label{ST4p}
\{Q_\mu,Q_\nu\}=(\gamma^m)_{\mu\nu}P_m\,.
\ee
Though  both algebras have the same number of the supercharges $Q_\mu$, their anti--commutator closes on different sets of bosonic generators. Thus, the super--Poincar\'e algebra \eqref{ST4p} is not a subalgebra of \eqref{ST4}. Hence the representations of \eqref{ST4} do not split into (finite--dimensional) representations of the standard super--Poincar\'e algebra. In this sense the supersymmetric higher--spin systems under consideration differ from  most of supersymmetric models of finite--dimensional super--Poincar\'e or AdS higher--spin supermultiplets  considered in the literature (see e.g. \cite{Curtright:1979uz,Vasiliev:1980as,Bellon:1986ki,Fradkin:1987ah,Bergshoeff:1988jm,Konstein:1989ij,
Kuzenko:1993jp,Kuzenko:1993jq,Kuzenko:1994dm,Buchbinder:1995ez,Gates:1996my,Gates:1996xs,Sezgin:1998gg,
Alkalaev:2002rq,Engquist:2002gy,Sezgin:2012ag,Zinoviev:2007js,Fotopoulos:2008ka,Gates:2013rka,Gates:2013ska,
Candu:2014yva,Kuzenko:2016bnv, Buchbinder:2016jgk, Kuzenko:2016qwo, Buchbinder:2017izy, Buchbinder:2017nuc}).

\section{Hyperspace extension of supersymmetric AdS spaces} \label{HEADS} \setcounter{equation}0

In Section \ref{SectionAdS} we have seen  that  the hyperspace extension
of AdS spaces are $Sp(n)$ group manifolds. In this section we consider their minimal supersymmetric extension, namely
 $OSp(1|n)$ supergroup manifolds.

The $OSp(1|n)$ superalgebra is formed by $n$ anti--commuting supercharges ${\mathcal Q}_\alpha$ and $\frac{n(n+1)}{2}$ generators $M_{\alpha\beta}=M_{\beta\alpha}$ of $Sp(n)$
\bea\label{OSp}
&\{{\mathcal Q}_\alpha,{\mathcal Q}_\beta\}=2M_{\alpha\beta}\,,\qquad [{\mathcal Q}_\alpha,M_{\beta\gamma}]=\frac{i\xi}2C_{\alpha(\beta}\,{\mathcal Q}_{\gamma)},&\nonumber\\
&[M_{\alpha\beta},M_{\gamma\delta}]=-\frac{i\xi}2(C_{\gamma(\alpha}M_{\beta)\delta}+C_{\delta(\alpha}M_{\beta)\gamma})\,,&
\eea
\if{}
where $C_{\alpha\beta}=-C_{\beta\alpha}$ is the $Sp(n)$ invariant symplectic metric and $\xi$ is a parameter of inverse dimension of length related to the $AdS$ radius via $r=2/\xi$.
\fi
The $OSp(1|n)$ algebra \eqref{OSp} is recognized as a subalgebra of  $OSp(1|2n)$ (see  the subsestion \ref{fss}) with the identifications
\be\label{generators}
{\mathcal Q}_\alpha=(Q_\alpha+\frac\xi 4 S_\alpha), \qquad M_{\alpha\beta}=P_{\alpha\beta}-\frac {\xi^2}{16}K_{\alpha\beta}-\frac \xi 4G_{(\alpha\beta)}\,.
\ee
\if{}
where $S_\alpha=S^\beta C_{\beta\alpha}$, $G_{\alpha\beta}=G_\alpha{}^\gamma C_{\gamma\beta}$ and $K_{\alpha\beta}=K^{\gamma\delta}C_{\gamma\alpha}C_{\delta\beta}$.
\fi

The $OSp(1|n)$ manifold is parameterized by the coordinates $(X^{\mu\nu},\theta^\mu)$ and its geometry is described by the Cartan forms
\be\label{Cartan}
\Omega={\mathcal O}^{-1}d{\mathcal O}(X,\theta)= -i\Omega^{\alpha\beta}M_{\alpha\beta}+iE^\alpha{\mathcal Q}_\alpha\,,
\ee
where ${\mathcal O}(X,\theta)$ is an $OSp(1|n)$ supergroup element. The Cartan forms satisfy the Maurer--Cartan equations associated with the $OSp(1|n)$ superalgebra \eqref{OSp}
\be\label{MC-11}
d\Omega^{\alpha\beta}+\frac \xi 2\Omega^{\alpha\gamma}\wedge \Omega_{\gamma}{}^{\beta}=-i E^\alpha \wedge E^\beta, \qquad dE^\alpha+\frac \xi 2E^{\gamma}\wedge \Omega_{\gamma}{}^{\alpha}=0\,,
\ee
with the external differential acting from the right.

\subsection{GL flatness of $OSp(1|n)$ group manifolds}

There is a supersymmetric generalization of the $GL(n)$ flatness property of $Sp(n)$ group manifolds to the case
of $OSp(1|n)$ supergroup manifolds \cite{Plyushchay:2003gv}. In particular,
the Maurer-Cartan equations \eqref{MC-11} are solved by the following forms
\be\label{Omega}
\Omega^{\alpha \beta} = d X^{\mu \nu} G_\mu{}^\alpha G_\nu{}^\beta(X) + \frac{i}{2} (\Theta^\alpha {\mathcal D} \Theta^\beta +
\Theta^\beta {\mathcal D} \Theta^\alpha) = \Pi^{\mu\nu} \,{\mathcal G}_\mu{}^\alpha \, {\mathcal G}_{\nu}{}^{\beta}(X,\Theta),
\ee
\begin{eqnarray}\label{Ealpha}
E^\alpha = P(\Theta^2) {\mathcal D} \Theta^\alpha - \Theta^\alpha{\mathcal D} P(\Theta^2)\,
\end{eqnarray}
where $\Theta$ is related to $\theta$ as follows
\begin{equation}\label{Theta}
\theta^\alpha=\Theta^\beta G_{\beta}^{-1\alpha}P^{-1}(\Theta^2),\qquad \Theta^2=\Theta^\alpha\Theta_\alpha,\qquad P^2(\Theta^2)=1+\frac{i\xi}8\Theta^2\,,
\end{equation}
while the covariant derivative
\be\label{mathcalD}
{\mathcal D}\Theta^\alpha=d\Theta^\alpha+\frac\xi 4 \Theta^\beta\,\omega_\beta{}^\alpha(X)\,,
\ee
contains the Cartan form of the $Sp(n)$ group manifold
\be\label{omega}
\omega^{\alpha\beta}(X)=dX^{\mu\nu}G_\mu{}^\alpha(X) G_{\nu}{}^\beta(X),
\ee
and
\begin{equation}\label{calG}
{\mathcal G}_{\alpha}{}^\beta(X,\Theta)=G_{\alpha}{}^\beta(X)-\frac{i\xi}8(\Theta_\alpha-2G_{\alpha}{}^\gamma (X)
\Theta_\gamma)\Theta^\beta,
\end{equation}
where $G_{\alpha}{}^\beta(X)$ is given in \p{sol3}.
The inverse matrix of (\ref{calG}) is
\begin{eqnarray}\label{calGn-or1}
{\mathcal G}_{\alpha}^{-1\beta}(X,\Theta)&=&G_{\alpha}^{-1\beta}(X)
-\frac{i\xi}{8}(\Theta^\delta G_{\delta\alpha}^{-1}(X))\,(\Theta^\delta\,G_{\delta}^{-1\beta}(X))P^{-2}(\Theta^2)\nonumber\\
&=&G_{\alpha}^{-1\beta}(X)
-\frac{i\xi}{8}\theta_\alpha\,\theta^\beta\,=\delta_\alpha^\beta+\frac\xi 4(X_\alpha{}^\beta-\frac i2 \theta_\alpha\,\theta^\beta)
\end{eqnarray}
with $G_{\alpha}^{-1\beta}(X)$ given in \p{sol2}.

\subsection{Field equations on $OSp(1|n)$ supergroup manifold}

The scalar superfield equation on $OSp(1|n)$ has the form \cite{Bandos:2004nn}
\be\label{Osp}
\left({\nabla}_{[\alpha }{\nabla}_{\beta ]}-\frac{i\xi}8C_{\alpha \beta }\right)\Phi_{OSp}(X,\theta)=0\,,
\ee
where the Grassmann--odd covariant derivatives $\nabla_\alpha$ and their bosonic counterparts $\nabla_{\alpha\beta}$ satisfy the $OSp(1|n)$ superalgebra similar to \eqref{OSp}, namely
\be\label{antic}
\{{\nabla}_{\alpha},{\nabla}_{\beta}\}=2i \nabla_{\alpha\beta}\,
\ee
\be\label{vs}
[\nabla_\gamma,\nabla_{\alpha\beta}]=\frac{\xi}2C_{\gamma(\alpha}\,\nabla_{\beta)},
\ee
\begin{equation} \label{algebra}
[\nabla_{\alpha \beta}, \nabla_{\gamma \delta}] = \frac{\xi}{2}(C_{\alpha (\gamma} \nabla_{\delta)\beta}+ C_{\beta (\gamma} \nabla_{\delta)\alpha }
)\,.
\end{equation}
while the $OSp(1|n)$ covariant derivatives are obtained from the flat superspace ones by the following GL transformations
\bea\label{nablaD}
\nabla_\alpha&=&{\mathcal G}_{\alpha}^{-1\,\mu}(X,\Theta)\,D_\mu\,, \\ \nonumber
\nabla_{\alpha\beta}&=&{\mathcal G}_{\alpha}^{-1\,\mu}(X,\Theta)\,
{\mathcal G}_{\beta}^{-1\nu} (X,\Theta)\left(\partial_{\mu\nu}+2i{D}_{(\mu}\ln\left((\det G (X))^{\frac 12}P^{-1}(\Theta^2)\right)\,{D}_{\nu)}\right).
\eea

\subsubsection{Connection between superfields on flat Hyper-Superspace and on $OSp(1|n)$ supergroup manifolds}
Using the relations given in Appendix \ref{CC} one can show that the superfield $\Phi_{OSp}(X,\theta)$ satisfying \eqref{Osp}
is related to the superfield $\Phi(X,\theta)$ satisfying the flat superspace equation \eqref{DDPhi} by the super--Weyl transformation
\begin{eqnarray}\label{W}
&&\Phi_{OSp(1|n)}(X,\theta)=({\det {\mathcal G}(X, \Theta)})^{-\frac 12}\,\Phi_{flat}(X,\theta) = \\ \nonumber
&&=({\det {G}(X)})^{-\frac 12}P(\Theta^2)\,\Phi_{flat}(X,\theta),
\end{eqnarray}

Substituting \eqref{bftheta} into \eqref{W} and using the definition (\ref{Theta}), together with the fact that on the mass shell all higher components in \eqref{bftheta} vanish, we find
\begin{eqnarray}\label{composp} \nonumber
&&\Phi_{OSp(n)}(X,\theta)=({\det {G}(X)})^{-\frac 12}\,b(X)+ \\
&&+\Theta^\alpha ({\det {G}(X)})^{-\frac 12}\,G_\alpha^{-1\mu}(X) \,f_\mu(X) +O(\Theta^2,b(X)),
\end{eqnarray}
where the first two terms are the fields
\be\label{BF1}
B(X)=({\det {G}}(X))^{-\frac 12}\,b(X),\qquad F_\alpha(X)=({\det {G}(X)})^{-\frac 12}\,G_\alpha^{-1\mu}(X) \,f_\mu(X)
\ee
propagating on the $Sp(n)$ group manifold, and $O(\Theta^2,b(X))$ stands for higher order terms in $\Theta^2$ which only depend on $b(X)$. The fields \eqref{BF1} satisfy the equations of motion on $Sp(n)$ group manifolds \p{bsp}--\p{fsp}.
 Note that in these equations the covariant derivatives are restricted to the bosonic group manifold $Sp(n)$, \emph{i.e.} $\nabla_{\alpha\beta}=G^{-1\,\mu}_\alpha(X)\,G^{-1\,\nu}_\beta(X)\,\partial_{\mu\nu}$.

\subsection{$OSp(1|2n)$ transformations of superfields}

Since the flat superspace field equation is invariant under the generalized superconformal $OSp(1|2n)$ transformations \eqref{deltaPhi}, the above relation leads us to conclude that also the $OSp(1|n)$ superspace equations \eqref{Osp} are invariant under the $OSp(1|2n)$ transformations, under which the superfield $\Phi_{OSp}(X,\theta)$ varies as follows
\bea\label{deltaPhi-osp}
\delta\Phi_{OSp}(X, \theta) &=&-(\epsilon^\mu\,{\mathbb Q}_\mu+\xi_\mu\, {\cal S}^\mu+ia^{\mu\nu}\,{\cal P}_{\mu\nu}+ik_{\mu\nu}\,{\cal K}^{\mu\nu}+ig_\mu{}^\nu\,{\cal G}_{\nu}{}^\mu)\,\Phi_{OSp} (X, \theta)\nonumber\\
&&-\frac 12\, \left(g_\mu{}^\mu-
 k_{\mu \nu} (X^{\mu \nu}+\frac i2 \theta^\mu\theta^\nu)+\xi_\mu\,\theta^\mu\right)\,\Phi_{OSp} (X, \theta)\,.
\eea
Here,
\be\label{KP}
{\cal P}_{\mu \nu}= - i{\cal D}_{\mu \nu} = -i (\partial_{\mu \nu} + \frac{\xi}{8} {\mathcal G}_{(\mu \nu)} (X, \Theta) )\,,
\ee
and
\be\label{KQ}
{\mathbb Q}_\mu= Q_\mu - \frac{i \xi}{8} \Theta_\mu P(\Theta)\,.
\ee
Using the relations given in the Appendix \ref{CC}
one may check that
the operators (\ref{KP}) and (\ref{KQ}) obey the flat hyperspace supersymmetry algebra
\be \label{KALG}
[{\cal P}_{\mu \nu}, {\cal P}_{\rho \sigma}] =0, \qquad \{ {\mathbb Q}_{\mu }, {\mathbb Q}_{\nu}  \} =- 2 {\cal P}_{\mu \nu},
\qquad [{\cal P}_{\mu \nu}, {\mathbb Q}_{\rho }] =0 \,.
\ee
The other generators of the $OSp(1|2n)$ are
\be\label{KS}
{\cal S}^\mu=-(X^{\mu\nu}+\frac i2\theta^\mu\theta^\nu){\mathbb Q}_\nu\,, \quad
{\cal G}_{\mu}{}^{\nu}=-{2i}(X^{\nu\rho}+\frac i2 \theta^\nu\theta^\rho){\cal D}_{\rho\mu}-i\theta^\nu\,{\mathbb Q}_\mu\,,
\ee
and
\be\label{KK}
{\cal K}^{\mu\nu}=i(X^{\mu\rho}+\frac i2\theta^\mu\theta^\rho)(X^{\nu\lambda}+\frac i2\theta^\nu\theta^\lambda){\cal D}_{\rho\lambda}-i\theta^{(\mu}{\cal S}^{\nu)}\,.
\ee
Taking into account  the commutation relations (\ref{KALG}) we see that
the operators ${\mathbb Q}_\mu, {\cal S}^\mu, {\cal P}_{\mu \nu}, {\cal G}_{\mu}{}^{\nu}$ and ${\cal K}^{\mu\nu} $
obey the same $OSp(1|2n)$ algebra  as the operators
${ Q}_\mu$, ${ S}^\mu,$ $ { P}_{\mu \nu}$, ${G}_{\mu}{}^{\nu}$ and $ K^{\mu\nu} $
considered in the subsection \ref{fss}.

\section{Generalized CFT. Part I. Correlation functions in $OSp(1|2n)$--Invariant models } \label{s5} \setcounter{equation}0

In the previous sections, we have described the generalized conformal group $Sp(2n)$ and generalized conformal supergroup
$OSp(1|2n)$. We introduced the fundamental fields and superfields and showed how they transform
under generalized conformal transformations.

In this Section we shall construct two-, three- and four-point correlation functions of these fields, by requiring
the $Sp(2n)$ symmetry of the correlators, i.e. by solving the corresponding Ward identities.
In other words we will follow the conventional approach adopted in multidimensional CFTs (see e.g., \cite{Osborn:1993cr}).
In particular, we will consider $OSp(1|2n)$ invariant correlation functions from which the $Sp(2n)$
invariant correlation functions can be recovered
as components of the expansions of the former in series of the Grassman coordinates $\theta^\mu$. $Sp(2n)$-invariant correlation functions in the tensorial spaces have been studied in \cite{Vasiliev:2003jc,Florakis:2014kfa,Florakis:2014aaa,Skvortsov:2016lbh} and in the unfolded formulation in \cite{Didenko:2012tv}.

\subsection{Two--Point functions}

Let us denote the two-point correlation function by
\be
W(Z_1,Z_2)= \langle \Phi(X_1, \theta_1)  \Phi(X_2, \theta_2)   \rangle \,.
\ee
The invariance under supersymmetry transformation generated by the operators $Q$, eq. \eqref{PQ}, requires that
\be
\epsilon^\mu \left( \frac{\partial}{\partial \theta_1^\mu}  - i \theta_1^\nu \frac{\partial }{\partial X_1^{\mu \nu}}
+
\frac{\partial}{\partial \theta_2^\mu}  - i \theta_2^\nu\frac{\partial }{\partial X_2^{\mu \nu}}
\right)W(Z_1, Z_2)=0 \,,
\ee
which implies
\be
 \langle \Phi(X_1, \theta_1)  \Phi(X_2, \theta_2)   \rangle= W({\rm det} |Z_{12}|),
\ee
where
\be
Z_{12}^{\mu \nu} = X_1^{\mu \nu} - X_2^{\mu \nu} - \frac{i}{2} \theta_1^\mu \theta_2^\nu -
\frac{i}{2} \theta_1^\nu \theta_2^\mu \,
\ee
is the interval between two points in hyper--superspace which is invariant under the rigid supersymmetry transformations \eqref{susyv}.

We next require the invariance of the correlator under the $S$-supersymmetry \eqref{S-1}
\bea \nonumber
&&\xi_\mu \left[  ( X_1^{\mu \nu} + \frac{i}{2} \theta_1^\mu \theta_1^\nu )
 \left ( \frac{\partial}{\partial \theta_1^\nu}  - i \theta_1^\rho \frac{\partial }{\partial X_1^{\nu \rho}} \right ) +
( X_2^{\mu \nu} + \frac{i}{2} \theta_2^\mu \theta_2^\nu )
 \left ( \frac{\partial}{\partial \theta_2^\nu}  - i \theta_2^\rho \frac{\partial }{\partial X_2^{\nu \rho }} \right ) \right ]
 \cdot \\ \nonumber
 && \cdot W({\rm det} |Z_{12}|) + \\
&& + \xi_\mu \left (  \frac{i}{2} \theta^\mu_1 + \frac{i}{2} \theta^\mu_2 \right)  W({\rm det} |Z_{12}|)=0 \,,
\eea
which is solved by
\be \label{2PTF}
 W({\rm det} |Z_{12}|)= c_2 ({\rm det} |Z_{12}|)^{-\frac{1}{2}} \quad \Rightarrow \quad \langle \Phi(X_1, \theta_1)  \Phi(X_2, \theta_2)   \rangle= c_2
 ({\rm det} |Z_{12}|)^{-\frac{1}{2}}\,.
\ee
The two--point function (\ref{2PTF})   reproduces
the correlators of the component bosonic and fermionic hyperfields $b(X)$ and $f_\mu(X)$
after the expansion of the former in powers of the Grassmann coordinates $\theta_1^{( \mu} \theta_2^{\nu )}$.
Since on the mass shell the superfield (\ref{bftheta}) has only two non--zero components,
all terms  in the $\theta$-expansion  of the two-point function  (\ref{2PTF}),
starting from the ones quadratic  in  $\theta_1^{( \mu} \theta_2^{\nu )}$, should vanish.
This is indeed the case, as a consequence of the field equations.

To see this,  let us recall that in the separated points the two--point function of the bosonic hyperfield of weight $\frac 12$ satisfies the free field
equation. Therefore for $X^1_{\alpha \beta} \neq X^2_{\alpha \beta}$ one has\footnote{When the two points coincide, one can define an analog
of the Dirac delta-function in the tensorial spaces, see \cite{Vasiliev:2001dc} for the relevant discussion.}
 \be \label{2ptch}
(\partial^1_{\mu \nu} \partial^1_{\rho \sigma }  - \partial^1_{\mu \rho} \partial^1_{\nu \sigma } )
\langle b(X_1) b(X_2) \rangle=
(\partial^1_{\mu \nu} \partial^1_{\rho \sigma }  - \partial^1_{\mu \rho} \partial^1_{\nu \sigma } )
({\rm det} |X_{12}|)^{-\frac{1}{2}}=0 \,.
\ee
Similarly, for $X^1_{\alpha \beta} \neq X^2_{\alpha \beta}$ the fermionic two--point function satisfies the free field equation for the fermionic hyperfield.
Written in terms of  the superfields,  these equations are encoded in the superfield equation (for $Z_{12}\not =0$)
\be\label{EOM2}
(D^1_{\mu } D^1_{\nu }  - D^1_{\nu} D^1_{\mu } )
\langle \Phi(X_1, \theta_1)  \Phi(X_2, \theta_2) \rangle=
(D^1_{\mu } D^1_{\nu }  - D^1_{\nu} D^1_{\mu } )
({\rm det} |Z_{12}|)^{-\frac{1}{2}}=0.
\ee
Expanding the two--point function $({\rm det} |Z_{12}|)^{-\frac{1}{2}}$ in powers of
the Grassmann variables
\bea \label{expansion}
&&({\rm det} |Z_{12}|)^{-\frac{1}{2}}
=({\rm det} |X_{12}|)^{-\frac{1}{2}} - \\ \nonumber
&&-i \partial_{\alpha \beta}  ({\rm det} |X_{12}|)^{-\frac{1}{2}}
\theta_1^{( \alpha} \theta_2^{\beta )} - \frac{1}{2}
\partial_{\gamma \delta}
\partial_{\alpha \beta}  ({\rm det} |X_{12}|)^{-\frac{1}{2}}\theta_1^{( \alpha} \theta_2^{\beta )}\theta_1^{( \gamma} \theta_2^{\delta )}
+ \ldots\,,
\eea
 one may see that the terms in the expansion starting from
$(\theta_1^{( \mu} \theta_2^{\nu )})^2$ vanish due to the free field equation
(\ref{2ptch}).
From equations   (\ref{2PTF}),   (\ref{expansion}) and from the explicit form of the superfield
(\ref{bftheta}), one may immediately reproduce the correlation functions for the component fields
\cite{Vasiliev:2003jc}
\be
\langle b(X_1) b(X_2) \rangle = c_2 ({\rm det} |X_{12}|)^{-\frac{1}{2}}\,,
\ee
\be
\langle f_\mu(X_1) f_\nu (X_2) \rangle = \frac{ic_2}{2} (X_{12})^{-1}_{\mu \nu} ({\rm det} |X_{12}|)^{-\frac{1}{2}}\,.
\ee

The two-point functions on the $OSp(1|n)$ manifold may now be obtained from
(\ref{2PTF}) via the rescaling \eqref{W}, which relates the superfields in flat superspace and on the $OSp(1|n)$ group manifold
\bea\label{osp2p}
&&\langle \Phi_{OSp}(X_1, \theta_1)  \Phi_{OSp}(X_2, \theta_2)   \rangle = \\ \nonumber
&& ({\rm det}\, G (X_1))^{-\frac{1}{2}} P(\Theta_1^2) ({\rm det}\, G (X_2))^{-\frac{1}{2}} P(\Theta_2^2)
\langle \Phi(X_1, \theta_1)  \Phi(X_2, \theta_2)\rangle\,.
\eea
Finally, as in the $D=3$ case, one may derive the superconformally invariant two--point function for superfields carrying an arbitrary generalized conformal weight $\Delta$, which on flat hyper superspace has the form
\be\label{2pd}
\langle \Phi^{\Delta_1}(X_1, \theta_1)  \Phi^{\Delta_2}(X_2, \theta_2)   \rangle= c_2
 ({\rm det} |Z_{12}|)^{-\Delta}\,, \qquad \Delta_1=\Delta_2=\Delta\,.
 \ee

\subsection{Three--Point functions} \label{susy3d}
The three--point functions for the superfields with arbitrary generalized conformal dimensions $\Delta_i$, $(i=1,2,3)$
\be
W(Z_1,Z_2, Z_3)= \langle \Phi(X_1, \theta_1) \Phi(X_2, \theta_2)   \Phi(X_3, \theta_3) \rangle \,,
\ee
may be computed in a way similar to the two--point functions using the superconformal Ward identities.
The invariance under $Q$--supersymmetry implies that they depend on the superinvariant intervals $Z_{ij}$,
\emph{i.e.}
\be
 \langle \Phi(X_1, \theta_1)  \Phi(X_2, \theta_2)   \Phi(X_3, \theta_3) \rangle= W(Z_{12}, Z_{23}, Z_{31})\,,
\ee
where
\be\label{differ}
Z_{ij}^{\mu \nu} = X_i^{\mu \nu} - X_j^{\mu \nu} - \frac{i}{2} (\theta_i^\mu \theta_j^\nu +\theta_i^\nu \theta_j^\mu)\,, \qquad i,j=1,2,3\,.
\ee
Invariance under $S$--supersymmetry then fixes  the form
of the function $W$ to be
\bea \label{3PTS}
&\langle \Phi(X_1, \theta_1) \Phi(X_2, \theta_2)   \Phi(X_3, \theta_3) \rangle = \\ \nonumber
&=c_3(\det Z_{12})^{-\frac{1}{2}(\Delta_1+\Delta_2 -\Delta_3)} (\det Z_{23})^{-\frac{1}{2}(\Delta_2+\Delta_3 -\Delta_1)}
 (\det Z_{31})^{-\frac{1}{2}(\Delta_3+\Delta_1 -\Delta_2)} \,.
\eea
Let us note that the three--point function is not annihilated by the operator entering the free equations of motion (\ref{DDPhi})
for generic values of the generalized conformal dimensions, including the case in which the values
of all the generalized conformal dimensions are canonical
 \bea
&&(D^1_{\mu } D^1_{\nu }  - D^1_{\nu} D^1_{\mu } )
\langle \Phi(X_1, \theta_1), \Phi(X_2, \theta_2) , \Phi(X_2, \theta_2)    \rangle= \\ \nonumber
&&=c_3(D^1_{\mu } D^1_{\nu }  - D^1_{\nu} D^1_{\mu } ) \left (
({\rm det} |Z_{12}|)^{-\frac{1}{4}}({\rm det} |Z_{23}|)^{-\frac{1}{4}}   ({\rm det} |Z_{31}|)^{-\frac{1}{4}}  \right ) \neq 0 \,.
\eea

Again, the three--point functions on the supergroup manifold $OSp(1|n)$ can be obtained via the Weyl rescaling
(\ref{W}), as in the case of the two--point functions \eqref{osp2p}
\bea\label{osp2p3}
&&\langle \Phi_{OSp}(X_1, \theta_1)  \Phi_{OSp}(X_2, \theta_2) \Phi_{OSp}(X_3, \theta_3)  \rangle = \\ \nonumber
&&= ({\rm det}\, G (X_1))^{-\frac{1}{2}} P(\Theta_1^2) ({\rm det}\, G (X_2))^{-\frac{1}{2}} P(\Theta_2^2)
({\rm det}\, G (X_3))^{-\frac{1}{2}} P(\Theta_3^2) \cdot \\ \nonumber
&&\cdot \langle \Phi(X_1, \theta_1)  \Phi(X_2, \theta_2)  \Phi(X_3, \theta_3) \rangle\,.
\eea

\subsection{Four--Point functions}

Finally, let us consider, first in flat hyper superspace, the correlation function of four real scalar superfields
with arbitrary generalized conformal dimensions,
 $\Delta_i$ ($i=1,2,3,4$)
\be
W(Z_1,Z_2, Z_3)= \langle \Phi(X_1, \theta_1) \Phi(X_2, \theta_2)   \Phi(X_3, \theta_3)  \Phi(X_4, \theta_4)\rangle\,.
\ee
Invariance under $Q$--supersymmetry again implies that
the correlation function depends only on the superinvariant intervals
$Z_{ij}^{\mu \nu}$ \eqref{differ}. Following  the analogy with conventional conformal field theory we find
\begin{equation}\label{4PF}
W(X_1, X_2, X_3, X_4) = c_4\,\prod_{ij, i<j}\frac{1}{{(\det |Z_{ij}|)}^{k_{ij}}   }
{\tilde W} \left (  z,z' \right)\,,
\end{equation}
with $W$ being an arbitrary function of the  cross-ratios
\begin{equation}
	z= \det\left(\frac{|Z_{12}||Z_{34}|}{|Z_{13}||Z_{24}|}\right)~,\qquad z'= \det\left(\frac{|Z_{12}||Z_{34}|}{|Z_{23}||Z_{14}|}\right)\,,
\end{equation}
subject to the crossing symmetry constraints
\begin{equation}\label{crossingSym}
	 \tilde W(z,z')=\tilde W \left(\frac{1}{z},\frac{z'}{z}\right)=\tilde W \left(\frac{z}{z'},\frac{1}{z'}\right)\,.
\end{equation}
Furthermore, the $k_{ij}$'s are constrained by the invariance of the four--point function under the $S$--supersymmetry to satisfy
\begin{equation}
\sum_{j\neq i} k_{ij} =  \Delta_i \,.
\end{equation}
Similarly to the case of two-- and three--point functions, the four--point function
of the scalar superfields on $OSp(1|n)$ can be obtained from (\ref{4PF}) via  the Weyl re-scaling (\ref{W}).

\subsection{An Example.  ${\cal N}=1$ $D=3$ superconformal models} \label{example3d}

As we mentioned earlier, the case of $D=3$ is the simplest example of `hyperspace' which in this case coincides with the three-dimensional space time itself, and the fundamental fields are just the scalar $b(x)$ and the two-component spinor $f_\alpha(x)$.
All known results for three-dimensional (super)conformal theories are reproduced from the above generic formulas restricted to the case of $n=2$ and $D=3$, as we will show on the example of ${\cal N}=1$ $D=3$ superconformal two-- and three-point functions.

The superconformally invariant two- and three-point correlation functions of the ${\mathcal N}=1$, $D=3$ scalar supermultiplet
model have been constructed in \cite{Park:1999cw}.

Let us use the spinor--tensor representation for the description of the three--dimensional space--time coordinates
\be\label{x3}
x^{\alpha \beta}= x^{\beta \alpha} = x^m (\gamma_m)^{\alpha \beta},
\ee
where now $\alpha, \beta=1,2$ are $D=3$ spinorial indices and $m=0,1,2$ is the vectorial one. Since \eqref{x3} provides a representation of the symmetric $2 \times 2 $ matrices $x^{\alpha\beta}$, no extra coordinates, like $y^{mn}$, are present and, hence, no higher-spin fields.

\noindent
The inverse matrix  of \eqref{x3}, $x^{-1}_{\alpha\beta}$
\be
x^{\alpha \beta} \,x^{-1}_{\beta \gamma} = \delta_\alpha^\gamma\,,
\ee
takes the simple form
\be
x^{-1}_{\alpha \beta} = -\frac{1}{x^m x_m} x^n (\gamma_n)_{\alpha \beta} =- \frac{1}{x^2} x_{\alpha \beta}\,.
\ee
We may now consider a real scalar  superfield in $D=3$
\be \label{superfield}
\Phi(x, \theta) = \phi(x) + i \theta^\alpha f_\alpha (x) + \theta^\alpha \theta_\alpha F(x)\,,
\ee
with $ \phi(x)$ being a physical scalar, $f_\alpha(x)$  a physical fermion and $F(x)$ an auxiliary field.

If \eqref{superfield} satisfies the free equation of motion \eqref{DDPhi}, which in the $D=3$ case reduces to
\be\label{DDPhi3}
D^\alpha D_\alpha \Phi(x, \theta)=0\,.
\ee
This equation implies that on the mass shell the auxiliary field $F(x)$ vanishes, the scalar field $\phi(x)$ satisfies the massless Klein--Gordon equation and
$f_\alpha(x)$ satisfies the massless Dirac equation. The field equation \eqref{DDPhi3} is superconformally invariant if the superfield $\Phi(x,\theta)$ has the canonical conformal weight $\Delta=\frac 12$.

Let us consider a superconformal transformation of \eqref{superfield}.
The Poincar\'e supersymmetry transformations of $\Phi$ are
\be \label{PS}
\delta \Phi(x, \theta) = \epsilon^\alpha \left( \frac{\partial}{ \partial \theta^\alpha} - i  \theta^{ \beta}\frac{\partial}{ \partial x^{\alpha \beta}} \right) \Phi(x, \theta) =  \epsilon^\alpha  Q_\alpha \Phi(x, \theta)\,.
\ee
They encode the supersymmetry transformations of the component fields
\begin{eqnarray}
&&\delta \phi (x) = i \epsilon^\alpha f_\alpha(x)\,, \\
&&\delta f_\alpha(x) = -2i \epsilon_\alpha F(x)  - \epsilon^\beta \partial_{\alpha \beta} \phi(x)\,, \\
&& \delta F(x) = \frac{1}{2} \epsilon^\alpha \partial_{\alpha \beta} f^\beta(x)\,,
\end{eqnarray}
where we have made use of the identity
\be
\theta^\alpha \theta^\beta =\frac{1}{2} C^{\alpha \beta} (\theta^\gamma \theta_\gamma)\,.
\ee
Under conformal supersymmetry, $\Phi(x, \theta)$ transforms as follows
\be \label{CS}
\delta \Phi(x, \theta)  = \xi_\alpha (x^{\alpha \beta} + \frac{i}{2} \theta^\alpha \theta^\beta) Q_\beta  \Phi(x, \theta)
- i (\xi_\alpha \theta^\alpha) \Delta \Phi(x, \theta) \,,
\ee
where $\Delta$ is the conformal weight of the superfield. The superconformal transformations of the component fields are
\begin{eqnarray}
&&\delta \phi(x)= i \xi_\alpha\,\, x^{\alpha \beta} f_\beta(x), \\
&& \delta f_\alpha(x)= -2i \xi_\beta \,\, x^\beta{}_\alpha F(x) + \xi_\beta \,\, x^{\beta \gamma}\,\, \partial_{\gamma \alpha} \phi(x)
+ \xi_\alpha \Delta \phi(x), \\
&& \delta F(x) = \frac{1}{2}\,\, \xi_\alpha \,\, x^{\alpha \beta} \partial_{\beta \gamma} f^\gamma(x) - \frac{1}{2} \xi_\alpha
\left( \frac{1}{2} - \Delta \right)  f^\alpha(x).
\end{eqnarray}
The conformal weights of $\phi$, $f_\alpha$ and $F$ are $\Delta$, $\Delta +\frac 12$ and $\Delta +1$, respectively.

As we have already  seen, the two-point function for a superfield of an arbitrary noncannonical dimension has the form \eqref{2pd}.
Expanding  the expression on the right hand side of \eqref{2pd} in powers of $\theta$, we obtain
\bea \label{expansion-2} \nonumber
({\rm det} |z_{12}|)^{-\Delta}&= ({\rm det} |x_{12}|)^{-\Delta} -i \partial_{\alpha \beta}  ({\rm det} |x_{12}|)^{-\Delta}\,
\theta_1^{( \alpha} \theta_2^{\beta )} \\
	&- \frac{1}{2}
\partial_{\gamma \delta}
\partial_{\alpha \beta}  ({\rm det} |x_{12}|)^{-\Delta}\,\theta_1^{( \alpha} \theta_2^{\beta )}\theta_1^{( \gamma} \theta_2^{\delta )} \,.
\eea
Using the identities
\be
\partial_{\alpha \beta}( {\rm det} |x|)^{-\Delta} = -\Delta\, x^{-1}_{\alpha \beta}\,\,\, {\rm det}| x|^{-\Delta} \,,
\ee
and
\be
\partial_{\alpha \beta} \partial_{\gamma \delta}  ({\rm det} |x|)^{-\Delta} =
\Delta \left( \Delta \,x^{-1 }_{\alpha \beta} x^{-1 }_{\gamma \delta}    + \frac{1}{2} x^{-1 }_{\alpha \gamma} x^{-1 }_{\beta \delta} +
\frac{1}{2} x^{-1 }_{\beta \gamma} x^{-1 }_{\alpha \delta} \right)
 ({\rm det} |x|)^{-\Delta} \,,
\ee
 one may rewrite  the expression  (\ref{expansion-2}) as
\bea \label{expansion-3-1}
(\det |z_{12}|)^{-{\Delta}}&=&
(\det |x_{12}|)^{-\Delta} \left(1 - {i \Delta} \frac{x^m_{12} (\gamma_m)_{\alpha \beta}}{x_{12}^2}
\theta_1^{\alpha} \theta_2^{\beta } - \frac{(2\Delta -1) \Delta}{4}
\frac{1}{ x_{12}^2} \theta_1^2 \theta_2^2 \right).\nonumber\\
&&
\eea
Thus, from  equations (\ref{expansion-2}) or \eqref{expansion-3-1}, one may immediately read off the expressions for the correlation functions of the component fields of the superfield
(\ref{superfield})
\be
\langle \phi(x_1) \phi(x_2) \rangle =c_2({\rm det} |x_{12}|)^{-\frac{1}{2}}\,,
\ee

\be
\langle f_\alpha(x_1) f_\beta(x_2) \rangle =- ic_2 \partial_{\alpha \beta}({\rm det} |x_{12}|)^{-\frac{1}{2}}\,,
\ee

\be\label{F}
\langle \phi(x_1) f_\alpha(x_2) \rangle =0\,, \qquad
 \langle F(x_1) \phi(x_2) \rangle =0\,, \qquad \langle F(x_1) f_\alpha(x_2) \rangle =0\,,
\ee

\be \label{FFa}
\langle F(x_1) F(x_2) \rangle = -\frac{c_2}{8}  \partial^{\alpha \beta } \partial_{\alpha \beta}  ({\rm det} |x|)^{-\Delta} \,.
\ee

Let us note that when the superfield $\Phi(x,\theta)$ has the canonical conformal dimension $\Delta=\frac{1}{2}$,
 due to the identity
\be
C^{\alpha \gamma} C^{\beta \delta} \partial^1_{\alpha \beta} \partial^1_{\gamma \delta }
({\rm det} |x_{12}|)^{-\frac{1}{2}}
= - \frac{1}{2} \eta^{mn}\frac\partial{\partial x_1^m}\frac\partial{\partial x_1^n} ({\rm det} |x_{12}|)^{-\frac{1}{2}},
\ee
the last term  in (\ref{expansion-2})
is proportional to the $\delta$--function if one moves to the Euclidean signature.
Then one has for the two--point function for the auxiliary field
\be \label{FF-1}
\langle F(x_1) F(x_2) \rangle = -\frac{\pi }{4} c_2 \delta^{(3)} (x_1-x_2).
\ee
Note that the correlation functions of the auxiliary field $F$ with the
 physical fields and with itself (for $x^m_1 \neq x^m_2$) vanish.

On the other hand, if the conformal weight of the superfield (\ref{superfield}) is anomalous, i.e.
$\Delta \neq \frac{1}{2}$, the correlators of the auxiliary field with the physical ones still vanish (in agreement with the fact that their conformal weights are different), but the $\langle F F \rangle $ correlator is
\bea\label{Fa}
\langle F(x_1) F(x_2) \rangle &=&- c_2 \frac{(2\Delta -1) \Delta}{4}
\frac{1}{ x_{12}^2}\,(\det |x_{12}|)^{-\Delta}= \\ \nonumber
&=&- c_2 \frac{(2\Delta -1) \Delta}{4}(\det |x_{12}|)^{-\Delta-1}.
\eea
This situation may correspond to an interacting quantum ${\mathcal N}=1$ superconformal field theory \cite{Synatschke:2010ub}, where the auxiliary field is non--zero, and fields acquire anomalous dimensions due to quantum corrections.

The consideration of three-point functions is analogous. Using the expression for the three-point function \p{3PTS}
and expanding it in series of the $\theta^\mu_i$ variables, we get for the component fields
 whose labels of  scaling dimension we skip for simplicity
\be \label{3p-21}
 \langle \phi(x_1)  \phi(x_2)   \phi(x_3) \rangle= c_3(\det |x_{12}|)^{-k_1} (\det |x_{23}|)^{-k_2} (\det |x_{31}|)^{-k_3}\,,
\ee
\bea \label{3p-3-1}
&& \langle f_\alpha(x_1) f_\beta (x_2)   \phi (x_3) \rangle =   \\ \nonumber
&&=- i c_3 \frac{ k_1 x^m_{12} (\gamma_m)_{\alpha \beta}}{ x^2_{12}}
  (\det |x_{12}|)^{-k_1} (\det| x_{23}|)^{-k_2} (\det |x_{31}|)^{-k_3}  \\ \nonumber
 &&  =- i c_3{ k_1 x^m_{12} (\gamma_m)_{\alpha \beta}}
  (\det |x_{12}|)^{-k_1-1} (\det| x_{23}|)^{-k_2} (\det |x_{31}|)^{-k_3}\,, \nonumber
\eea
\bea
&& \langle f_\alpha(x_1)    F(x_2)   f_\beta(x_3) \rangle= \\ \nonumber
&&=c_3
\frac{k_1 k_2}{2 x^2_{12} x^2_{23}} (\gamma_m)_\alpha{}^\delta (\gamma_n)_{\delta \beta} (x_{12}^m) (x_{23}^n)
(\det |x_{12}|)^{-k_1} (\det |x_{23}|)^{-k_2} (\det |x_{31}|)^{-k_3} \\ \nonumber
&&=c_3
\frac{k_1 k_2}{2 } (\gamma_m)_\alpha{}^\delta (\gamma_n)_{\delta \beta} (x_{12}^m) (x_{23}^n)
(\det |x_{12}|)^{-k_1-1} (\det |x_{23}|)^{-k_2-1} (\det |x_{31}|)^{-k_3}\,,
\eea
\be\label{3p-4-1}
 \langle F(x_1)  F(x_2)   \phi(x_3) \rangle =- \frac{c_3}{8}
\partial^m \partial_m((\det |x_{12}|)^{-k_1}) (\det| x_{23}|)^{-k_2} (\det |x_{31}|)^{-k_3}.
\ee

The remaining three-point functions  containing an odd number of fermions, as well as the correlator $\langle F\phi\phi\rangle$, vanish.
Note that, dimensional arguments would allow for a non--zero $\langle F\phi\phi\rangle$ correlator, but supersymmetry forces it to vanish. The correlator  $ \langle F(x_1)  F(x_2)   F(x_3) \rangle$ is zero as well, since it is proportional to
$(\gamma_m \gamma_n \gamma_p)x_{12}^m x_{23}^n x_{31}^p =2i \epsilon_{mnp}x_{12}^m x_{23}^n x_{31}^p=0.$

Moreover, from the above expressions we see that superconformal symmetry does not fix the values of the scaling dimensions $\Delta_i$.
This indicates that quantum operators may acquire anomalous dimensions and the quantum ${\mathcal N}=1$, $D=3$ superconformal theory of scalar superfields can be non-trivial, in agreement \emph{e.g.} with the results of \cite{Synatschke:2010ub}.

If the value of $\Delta$ were restricted by superconformal symmetry to its canonical value and no anomalous dimensions were allowed (for all the operators which are not protected by supersymmetry) one
would conclude that the conformal fixed point is that of the free theory. This is the case, for instance, for the ${\mathcal N}=1$, $D=4$ Wess-Zumino model in which the chirality of ${\mathcal N}=1$ matter multiplets and their three-point functions restricts the scaling dimensions of the chiral scalar supermultiplets to be canonical. This implies that in the conformal fixed point the coupling constant is zero, \emph{i.e.} the theory is free \cite{Ferrara:1974fv,Conlong:1993eu}.

\section{Generalized CFT. Part II} \label{sections6} \setcounter{equation}0
In this Section, we shall continue our consideration of the generalized CFT based on the
symmetries of the generalized conformal group $Sp(2n)$. We shall mainly follow \cite{Skvortsov:2016lbh}.

\subsection{Conserved currents}
In Section \ref{FH}, we introduced the bosonic and fermionic fields in hyperspace which play the role
of the scalar and fermionic fields in ordinary conformal field theory.
In order to continue the analogy with CFTs let us consider the fields
$ b^A_\Delta(X)$ and $f^A_{\mu \Delta}(X)$ where now $A=1.,,,.N$ is an index of an internal $O(N)$ group (not to be confused with the Weyl spinor indices of the previous Sections) and $\Delta$ are corresponding
generalized conformal weights.

The two point functions of these fields are similar to those obtained in the previous section, with an obvious generalization
including the ``color" indexes
\be \label{2ptbcolor}
\langle b^A_{\Delta_1}(X_1), b^B_{\Delta_2}(X_2)\rangle = c_{bb} (det|X_{12}|)^{-\Delta}  \,\,\,\, \delta^{AB},
\ee
\be \label{2ptfcolor}
\langle f^A_{\alpha (\Delta_1)}(X_1), f^B_{\beta (\Delta_2)}(X_2)\rangle = c_{ff} (det|X_{12}|)^{-\Delta}
 (X_{12})^{-1}_{\alpha \beta} \,\,\,\, \delta^{AB},
\ee
where ~$\Delta_1= \Delta_2 = \Delta$, ~and  ~$(X_{12})_{\alpha \beta}= (X_1)_{\alpha \beta} - (X_2)_{\alpha \beta}$.

~Having ~introduced ~global $O(N)$ ~symmetry, ~one ~can ~construct bosonic and  fermionic biliniears
\be \label{cu-1}
J_{\mu \nu}^{AB}(X) = b^A (X)\partial_{\mu \nu} b^B (X) -  b^B (X) \partial_{\mu \nu} b^A (X),
\ee
\be \label{cu-2}
J_{\mu \nu}^{AB} (X)= f_\mu^A (X) f_\nu^B (X) + f_\nu^A (X) f_\mu^B (X).
\ee
These bilinears correspond to conserved $O(N)$ currents. Indeed one can check that the currents
\p{cu-1} and \p{cu-2} satisfy the generalized conservation conditions (first introduced in \cite{Vasiliev:2002fs})
\be \label{gccj}
\partial_{\mu \nu} J_{\alpha \beta}^{AB} (X) -
\partial_{\mu \alpha} J_{\nu \beta}^{AB}(X)-
\partial_{\beta \nu} J_{\alpha \mu}^{AB}(X)+
\partial_{\beta \alpha} J_{\nu \mu}^{AB} (X)=0
\ee
provided that the fields $b^A(X)$ and $f^A_\mu(X)$ satisfy the free equations of motion
\p{b} and \p{f}.

Knowing the $Sp(2n)$ transformations \p{sp8fb}--\p{sp8ff} of the fields $b^A(X)$ and $f^A_\mu(X)$ and using the equations \p{cu-1} and \p{cu-2}, one can derive the $Sp(2n)$ transformations
of the conserved currents
\begin{eqnarray}
&&\delta_a J_{\mu \nu}^{AB}(X)= -a^{\alpha \beta} \partial_{\alpha \beta} J_{\mu \nu}^{AB} (X)   \\
&&\delta_g J_{\mu \nu}^{AB}(X) = - \left (    g_\alpha{}^\alpha + 2 g_\alpha{}^\beta X^{\alpha \gamma}
\partial_{\beta \gamma}\right ) J_{\mu \nu}^{AB} (X)
-2 g_{(\mu}{}^\rho J_{\rho \nu)}^{AB} (X) \label{BBBB} \\
&&\delta_k J_{\mu \nu}^{AB}(X) =  ( k_{\alpha \beta} X^{\alpha \beta} + k_{\alpha \beta}X^{\alpha \gamma}
X^{\beta \delta} \partial_{\gamma \delta} ) J_{\mu \nu}^{AB} (X)
+2k_{(\mu \alpha}X^{\alpha \beta} J_{\beta \nu)}^{AB} (X)
\end{eqnarray}
From this transformation laws i.e, from the coefficients in front of the terms $g_\alpha{}^\alpha $ and
$k_{\alpha \beta} X^{\alpha \beta} $ one can conclude that the generalized conformal dimension $\Delta_J$ of the currents
\p{cu-1} and \p{cu-2} is equal to $1$. The same conclusion can be reached from the fact that
\p{cu-1} and \p{cu-2} correspond to free currents and the  generalized conformal dimension of the fields
$b(X)$ and $f_\mu(X)$ is equal to $\frac{1}{2}$.
Using the general expression \p{DD}, one can see that the generalized conformal dimension
is related to the usual scaling dimension as follows.
Recall (see subsection \ref{glcg}) that  $SL(n)$ subalgebra of $GL(n)$ algebra is parameterized by
$l_\mu{}^\nu =g_\mu{}^\nu - \frac{1}{n}\delta_\mu^\nu g_\rho{}^\rho $.
Let us rewrite the equation \p{BBBB}  as
\be
\delta_g J_{\mu \nu}^{AB}(X) = - \left ( \frac{n+2}{n}   g_\alpha{}^\alpha + 2 g_\alpha{}^\beta X^{\alpha \gamma}
\partial_{\beta \gamma}\right ) J_{\mu \nu}^{AB} (X)
- 2l_{(\mu}{}^\rho J_{\rho \nu)}^{AB} (X)
\ee
and define a weight
$\Delta_1$ as follows
\be \label{Delta1}
\Delta_1 = 1 + \frac{2}{n}.
\ee
Then using the relations \p{DD} one can see that
\be
\Delta_{{ D},1} = D-1
\ee
which is the canonical conformal weight of a spin-$1$ field.

 \subsection{Stress tensor}

Since we are considering a generalized CFT it is natural to define a generalized stress tensor, which contains
a usual CFT stress tensor when projected to the $x$-subspace.
Taking
\be \label{st-1}
{\tilde T}_{\mu \nu, \rho \sigma}(X)= (\partial_{\mu \nu}b (X)) (\partial_{\rho \sigma}b (X)) - \frac{1}{3}
b(X)(\partial_{\mu \nu} \partial_{\rho \sigma}b (X))
\ee
and
\be \label{st-2}
{\tilde T}_{\mu \nu, \rho \sigma} (X)= f_\rho (X) \partial_{\mu \nu} f_\sigma (X)
\ee
we define the
 generalized stress tensor as a symmetrized combination
\be \label{st}
T_{\mu \nu, \rho \sigma} (X) = {\tilde T}_{\mu \nu, \rho \sigma}(X) + {\tilde T}_{\mu \rho, \nu \sigma}(X)
+{\tilde T}_{\mu \sigma, \nu \rho} (X)
\ee
The reason of taking the expression \p{st} as a definition for the generalized stress tensor instead of
\p{st-1} and \p{st-2} is that \p{st} transforms properly under the $Sp(2n)$ transformations
\be
\delta_a  T_{\mu \nu  \rho \sigma}(X)= -a^{\alpha \beta} \partial_{\alpha \beta} T_{\mu \nu, \rho \sigma}(X),
\ee
\begin{eqnarray}
\delta_g T_{\mu \nu \rho \sigma}(X)&=&  -( g_{\alpha}{}^{ \alpha}  + 2g_{\alpha \beta}X^{\alpha \gamma}
 \partial_{\beta \gamma} ) T_{\mu \nu \rho \sigma} (X) -\\ \nonumber
&&-g_{\mu}{}^{ \alpha}T_{\alpha \nu \rho \sigma}(X) -...
-g_{\sigma}{}^{ \alpha}T_{\mu \nu \rho \alpha}(X),
\end{eqnarray}
\begin{eqnarray}
\delta_k T_{\mu \nu \rho \sigma}(X)&=&  ( k_{\alpha \beta} X^{\alpha \beta} + k_{\alpha \beta}X^{\alpha \gamma}
X^{\beta \delta} \partial_{\gamma \delta} ) T_{\mu \nu \rho \sigma}(X) +\\ \nonumber
&&+k_{\mu \alpha}X^{\alpha \beta} T_{\beta \nu \rho \sigma}(X) +...
+k_{\sigma \alpha}X^{\alpha \beta} T_{\mu \nu \rho \beta}(X).
\end{eqnarray}
The transformations above are again derived using the transformations
for the free fields \p{sp8fb}--\p{sp8ff} and the explicit form of the
stress energy tensor \p{st}. Again, using \p {DD}, one can see  that the generalized conformal dimension
of the stress tensor is $\Delta_T=1$, whereas   the   conformal dimension $\Delta_2$ (analogous to the expression \p{Delta1}
for $s=1$ current)
is
\be
\Delta_2 = 1+ \frac{4}{n}
\ee
and the canonical spin-2 field weight is
$$
\Delta_{D,2}=D\,
$$
in compliance with the general formula $\Delta_{D,s}=D+s-2$.

Like the conserved current $J_{\mu \nu}^{AB}$, the stress energy tensor satisfies the generalized conservation conditions
\be \label{conservation-T}
\partial_{\mu \nu} T_{\alpha \beta \gamma \delta}(X)  -
\partial_{\mu \alpha} T_{\nu \beta \gamma \delta}(X)-
\partial_{\beta \nu} T_{\alpha \mu \gamma \delta}(X)+
\partial_{\beta \alpha} T_{\nu \mu \gamma \delta}(X)=0
\ee
provided the fields satisfy the free equations of motion  \p{b} and \p{f}.

\subsection{Higher spin conserved currents}
By analogy with $J_{\alpha\beta}(X)$ and $T_{\alpha\beta\gamma\delta}(X)$ one can introduce \cite{Vasiliev:2002fs} higher-spin conserved currents $T_{\alpha_1\ldots \alpha_{2s}}(X)$ ($2s=1,2,3,\ldots$)  which transform under $Sp(2n)$ as follows
\begin{equation} \label{TRT1s}
\delta_a T_{\alpha_1 \ldots \alpha_{2s}}(X)=- a^{\mu \nu} \partial_{\mu \nu} T_{\alpha_1 \ldots \alpha_{2s}}(X),
\end{equation}
\begin{eqnarray} \label{TRT2s}
\delta_g T_{\alpha_1 \ldots \alpha_{2s}}(X)&=&- ( \Delta_s\,g_\mu{}^\mu + 2 g_\nu{}^\mu X^{\nu \rho}
\partial_{\mu \rho})T_{\alpha_1 \ldots \alpha_{2s}}(X) - \\ \nonumber
&&-2sl_{(\alpha_1}{}^\mu T_{\alpha_2 \ldots \alpha_{2s})\mu }(X),
\end{eqnarray}
\begin{eqnarray} \label{TRT3s}
\delta_k T_{\alpha_1 \ldots \alpha_{2s}}(X)&=&( k_{\mu \nu}  X^{\mu \nu}+
 k_{\mu \nu}  X^{\mu \rho} X^{\nu \lambda}\partial_{\rho \lambda})
T_{\alpha_1 \ldots \alpha_{2s}}(X) + \\ \nonumber
&&+ 4 k_{\mu(\alpha_1 } X^{\mu \nu} T_{\alpha_2 \ldots \alpha_{2s})\nu}(X),
\end{eqnarray}
where
\be\label{Ds}
\Delta_s=1+\frac {2s}n\,.
\ee
Again, using the relations \p{DD}, one can see that
\be
\Delta_{{ D},s} = D+s-2
\ee
which is a conventional  expression for a canonical conformal weight for a field with spin $s$.

The higher spin currents obey $Sp(2n)$ conservation conditions \cite{Vasiliev:2002fs}
\begin{equation}\label{conservation-s}
\partial_{\mu \nu}T_{\alpha \beta \gamma(2s-2)}(X)-\partial_{\mu \alpha}T_{\nu \beta \gamma(2s-2)} (X)-
\partial_{\beta \nu}T_{\alpha \mu \gamma(2s-2)}(X) +
\partial_{ \alpha \beta}T_{\mu \nu \gamma(2s-2)}(X)=0.
\end{equation}

 \subsection{Two-point correlation functions of the currents}
We have already considered two-point functions for scalar and spinorial hyperfields \p{2ptbcolor}--\eqref{2ptfcolor}.
Using these expressions as well as the expressions
 for the generalized conserved currents \p{cu-1} -- \p{cu-2}, it
 is straightforward to compute the two--point functions of two currents
 \be \label{2ptj}
\langle   J^{AB}_{\alpha \beta}(X_1) ,J^{CD}_{\mu \nu}(X_2)    \rangle
 =C_{JJ} ({\det|X_{12}|})^{-1}  (P_{12})_{\alpha \beta, \mu \nu}(\delta^{AC} \delta^{BD}- \delta^{AD} \delta^{BC}).
\ee
Here, we introduced an $Sp(2n)$-invariant tensor structure\footnote{ When checking the invariance under the generalized conformal boosts
notice that
the first pair of the indices of $(P_{12})_{\alpha\beta,\gamma\delta}$ gets rotated with the matrix $k_{\alpha\sigma}X^{\sigma\delta}_1$ and the second pair gets rotated with $k_{\mu\sigma}X^{\sigma\delta}_2$.}
(which we call $P$--structure)
\be \label{deP}
(P_{ab})_{\alpha \beta, \mu \nu}= (X^{-1}_{ab})_{\mu \alpha} (X^{-1}_{ab})_{\nu \beta} + (X^{-1}_{ab})_{\nu \alpha} (X^{-1}_{ab})_{\mu \beta}
\ee
$a,b=1,2$ and $a\not= b$.
which will be one of the building blocks for higher point correlation functions as well.

One more building block for the correlation functions is
$(X_{12})_{\alpha \beta}^{-1}$ which is  $Sp(2n)$ invariant when considered as a bilocal tensor
\begin{eqnarray} \label{deX}
\delta_{tot} (X^{-1}_{12})_{\alpha\beta}&=&-(X^{-1}_{12})_{\alpha\gamma}(\delta X_1-\delta X_2)^{\gamma\delta}(X^{-1}_{12})_{\delta\beta}\nonumber\\
&+& 2g_{(\alpha}{}^\gamma (X^{-1}_{12})_{\beta)\gamma}k_{\alpha\gamma}X_1^{\gamma\delta} (X^{-1}_{12})_{\delta\beta}-(X^{-1}_{12})_{\alpha\delta}X_2^{\delta\gamma}k_{\gamma\beta}=0\,.\nonumber
\end{eqnarray}

Similarly, for the two stress tensors one finds
\be \label{2ptT}
\langle   T_{\alpha \beta \gamma \delta}(X_1) ,T_{\mu \nu \rho \sigma}(X_2)   \rangle=  C_{TT}\frac{1}{\det|X_{12}|}
 \left ( (P_{12})_{\alpha \beta, \mu \nu}(P_{12})_{\gamma \delta, \rho \sigma}+ symm. \right ),
\ee
where the total symmetrization of the both sets of indices $(\alpha \beta  \gamma \delta)$ and
$(\mu \nu  \rho \sigma)$ is assumed.

It is instructive to recall the similar expressions for two-point functions in the usual CFT
 \be
\langle   T^{(l)}_{\mu_1,..., \mu_n}(x_1) , T^{(l)}_{\nu_1,..., \nu_n}(x_2) \rangle = c_{TT}
\frac {g_{\mu_1 \nu_1}(x_{12})...g_{\mu_n \nu_n}(x_{12})}{(x_{12})^{l}} - traces
\ee
with
\be
g_{\mu \nu}= \delta_{\mu \nu}- \frac{x_\mu x_\nu}{x^2}.
\ee
Obviously, the  $Sp(2n)$-invariant  structure $(P_{12})_{\alpha \beta,\gamma \delta}$ is a generalization
of $g_{\mu \nu}$.  Notice also  that the expressions for two-point functions
\p{2ptj}--\p{2ptT} can be obtained from solving generalized Ward identities, as it has been done
for the case of scalar and spinor hyperfields.
The generalized Ward identity for an $n$--point function
\begin{equation}\label{multipoint}
\langle\Phi^{\Delta^{(1)}}_{\alpha_1\ldots \alpha_{r_1}}(X_1) \ldots \Phi^{\Delta^{(k)}}_{\beta_1\ldots \beta_{r_k}}(X_k)\rangle \equiv G_{\alpha_1\ldots \alpha_{r_1},\ldots,\beta_1 \ldots\beta_{r_k}}(X_1, \ldots,X_k)\,.
\end{equation}
is as follows
\begin{eqnarray}\label{deltaG}
&\sum_{i=1}^{k}\left [\Delta_i(g_\mu{}^{\mu}-k_{\mu\nu}X^{\mu\nu}_i)+\delta X^{\mu\nu}_i\frac{\partial}{\partial X^{\mu\nu}_i}\right]G_{\alpha_1\ldots \alpha_{r_1},\ldots,\beta_1 \ldots\beta_{r_k}}(X_1,\ldots,X_k)&\nonumber\\
&+\sum_{j=1}^{_1}(g_{\alpha_j}{}^{\mu_j}-k_{\alpha_{j}\nu}X_1^{\nu\mu_j})\,G_{\mu_1\ldots \mu_j\ldots \mu_{r_1},\ldots,\beta_1\ldots \beta_{r_k}}(X_1, \ldots,X_k)+\cdots &\nonumber\\
&+\sum_{j=1}^{r_k}(g_{\beta_j}{}^{\mu_j}-k_{\beta_{j}\nu}X_k^{\nu\mu_j})\,G_{\alpha_1\ldots \alpha_{r_k},\ldots,\mu_1\ldots \mu_j\ldots \mu_{r_k}}(X_1, \ldots,X_k) =0\,,&
\end{eqnarray}
It is straightforward to check that the two-point functions solve the equations \p{deltaG}.

\subsection{Three point functions: $bbb$ and $ffb$}
Three-point functions for three scalars and for two fermions  and a scalar (computed firstly in \cite{Vasiliev:2003jc}) have been given  in Section \ref{susy3d} in the supersymmetric form and as a particular example for $D=3$ were given in
Section \ref{example3d}.
The only difference with the case without supersymmetry is that the overall constants
in front of the non-supersymmetric ones are independent of each other
\be  \label{3pb}
 \langle   b_{\Delta_1}(X_1) b_{\Delta_2}(X_2) b_{\Delta_3}(X_3)    \rangle
= C_{bbb}\,{(\det |X_{12}|)}^{-{k_3}} \,{(\det |X_{23}|)}^{-{k_1}}\,{(\det |X_{13}|)}^{-{k_2}}\,,
\ee

\begin{equation} \label{2pfb}
 \langle   f_\alpha(X_1) f_\beta(X_2) b(X_3)    \rangle
= c_{ffb}\,(X^{-1}_{12})_{\alpha\beta}{(\det |X_{12}|)}^{-{k_3}} \,{(\det |X_{23}|)}^{-{k_1}}\,{(\det |X_{13}|)}^{-{k_2}}\,.
\end{equation}
\be \label{3k-1}
k_a =\frac{1}{2}( \Delta^{(a+1)} + \Delta^{(a+2)} - \Delta^{(a)}), \quad cycl. \quad (a=1,2,3).
\ee

 \subsection{Three-point functions with $J$ and $T$}
Now, we would like to consider three-point functions which include the generalized conserved current
$J^{AB}_{\alpha \beta}(X)$ and generalized stress tensor $T_{\alpha \beta \gamma \delta}(X)$.
These can give us an answer whether an interacting generalized
conformal field theory based on $Sp(2n)$ symmetry exists.
As we shall see below, the answer to this question is negative.

Our strategy is as follows. As we have seen the generalized conformal
weighs of $J^{AB}_{\alpha \beta}(X)$
and $T_{\alpha \beta \gamma \delta}(X)$ are equal to one, $\Delta_J= \Delta_T=1$. If we assume that the corresponding
symmetries are not broken by interactions, then the values of $\Delta_J$ and $\Delta_T$
will remain the same. Therefore, we would like to construct $Sp(2n)$-invariant three- and higher-order
correlation functions which include $J^{AB}_{\alpha \beta}(X)$,
$T_{\alpha \beta \gamma \delta}(X)$ and other operators ${\cal O}$
and see if the conservation conditions \p{gccj} and \p{conservation-T}
along with $Sp(2n)$ invariance allow for the operators
${\cal O}$ to have anomalous dimensions. We will find that this is unfortunately not the case for $n>2$.

First let us introduce one more $Sp(2n)$-invariant tensor structure (which we call $Q$--structure)
\be \label{Q}
(Q^c_{ab})_{\alpha \beta} = (X^{-1}_{ac})_{\alpha \beta} - (X^{-1}_{bc})_{\alpha \beta}, \quad a,b,c=1,2,3
\ee
This structure, along with \p{deP} and
\be \label{pab}
(p_{ab})_{\alpha \beta}=(X_a^{\alpha \beta} - X_b^{\alpha \beta})^{-1}, \quad a,b=1,2,  \quad a\not= b.
\ee
is a building block for all the $Sp(2n)$-invariant correlation functions. In other words,
the most general multi-point function can be written as a sum over all possible polynomials of a required rank of the three structures $p_{ab}=X^{-1}_{ab}$  , $P_{ab}$ and $Q^c_{ab}$   times a pre-factor
\be
\langle \Phi...\Phi\rangle = G(p_{ab},P_{ab},Q^c_{ab}|X_{ab}).
\ee

Following this prescription one can immediately write the  simplest three-point function of
two scalars (with  generalized conformal dimensions $\Delta_1= \Delta_2=\Delta$)
and a conserved current (with $\Delta_J=1$)
\begin{eqnarray}
&&\langle   b_{\Delta_1}(X_1) b_{\Delta_2}(X_2) J_{\alpha \beta}(X_3)    \rangle=  \\ \nonumber
&& =C_{bb J}
{(\det |X_{12}|)^{-k_3}  (\det |X_{13}|)^{-k_2} (\det |X_{23}|)^{-k_1}} (Q^3_{12})_{\alpha \beta},
 \nonumber
\end{eqnarray}
and a three-point function of the two scalars (with
$\Delta_1= \Delta_2=\Delta$)
and the stress tensor (with $\Delta_T=1$)
\begin{eqnarray} \label {bbT-TT}
&&\langle   b(X_1) b(X_2) T_{\alpha \beta \gamma \delta}(X_3)    \rangle
=C_{bb T}
{(\det |X_{12}|)}^{-k_3}  (\det |X_{13}|)^{-k_2} \times \\ \nonumber
 &&{(\det |X_{23}|)^{-k_1}}
 ((Q^3_{12})_{\alpha \beta}
(Q^3_{12})_{\gamma \delta}
+ (Q^3_{12})_{\alpha \gamma} (Q^3_{12})_{\beta \delta}
 + (Q^3_{12})_{\alpha \delta}
 (Q^3_{12})_{\beta \gamma}),
\end{eqnarray}
where $k_a$ are restricted according to \p{3k-1}.
One can see that $Sp(2n)$ invariance alone does not impose any requirement
on the generalized conformal dimension $\Delta$ of the scalar field.

The next step is to require the conservation of the current
$J^{AB}_{\alpha \beta}(X)$ and the stress tensor
$T_{\alpha \beta \gamma \delta}(X)$
according to the equations \p{gccj} and \p{conservation-T}. This implies
\be
k_1 = k_2 = \frac{1}{2}, \quad {\rm and~any} \quad k_3\,.
\ee
 Therefore, in this case, no restriction on generalized conformal dimension
 of the scalar field appears i.e.,  anomalous dimension and therefore interactions are allowed. At this, the current and the stress tensor remain conserved, and
 their dimensions remain canonical $\Delta_J= \Delta_T=1$.

The next nontrivial example is a three point-function of two conserved currents and one scalar
 operator ${\cal O}(X)$ of dimension $\Delta$.
From the $Sp(2n)$-invariance condition we have
\begin{align} \label{jjo}
 \langle   J_{\mu \nu }(X_1) \mathcal O(X_2) J_{\alpha \beta}(X_3)    \rangle =&{(\det |X_{12}|)^{-\frac \Delta 2}  (\det |X_{13}|)^{-\frac{2-\Delta}2} (\det |X_{23}|)^{-\frac\Delta 2}} \times \\
& \times \left( {\cal A} [ (Q^3_{12})_{\alpha \beta} (Q^1_{23})_{\mu\nu}]
+ {\cal B}(P_{13})_{ \mu\nu,\alpha\beta}\right) \nonumber
\end{align}
where ${\cal A}$ and $\cal B$ are some  constants.
Again, one can see that $Sp(2n)$ symmetry alone does not impose
any restriction on the generalized conformal dimension of  ${\cal O}(X)$.

However, imposing the current conservation condition \p{gccj}, one gets \be \label{A=B}
{\cal A} = {\cal B}, \quad and \quad \Delta=1\,,
\ee
that is the dimension of the operator ${\cal O}(X)$ is fixed
\footnote{Since the canonical dimension
of the field $b(X)$ is equal to $\frac{1}{2}$ it is natural to assume
that the operator ${\cal O}(X)$ is a composite one ${\cal} O(X) = b^2(X)$. }
by the current
conservation condition.

 Let us note that
 from the point of view
 of the $x$-space
the current $J^{AB}_{\alpha \beta}(X)$ contains higher spin currents as a result of its
expansion in series of $y$ coordinates.
Therefore, this result is in accordance with the theorem of \cite{Maldacena:2011jn} stating that the conformal field theories which contain conserved higher-spin currents should be free.

Let us note, however, that in the simplest case of $n=2$, i.e. $D=3$ CFTs with the $Sp(4)$ conformal group
the two conditions \p{A=B}
are reduced to one (see \cite{Skvortsov:2016lbh} for technical details)
\be
{\cal A}(D-1 - \Delta) - {\cal B} \Delta=0\,.
\ee
This means that
the conformal dimension $\Delta$ of the operator ${\cal O}(X)$
remains undetermined, and hence this analysis does not ban the existence of interacting $D=3$ CFTs, as is well known.

\subsection{General case}
Let us now discuss the general structure of the three-point correlators of conserved currents which are symmetric tensors of rank $r=2s$ with $s$ being an integer `spin'.
To this end, it is convenient to hide the tensor indices away by contracting them with auxiliary variables
$\lambda^\alpha_a$, where $a$ refers to the point of the operator insertion:
\begin{align}
(p_{ab})_{\alpha\beta}&\Rightarrow p_{ab}=(X^{-1}_{ab})_{\alpha\beta}\, \lambda^\alpha_a \lambda_b^\beta \quad \text{no summation over }a,b\,.\\
(P_{bc})_{\alpha\beta,\gamma\delta}&\Rightarrow P_{ab}= 2 p_{ab}p_{ba}=(P_{ab})_{\alpha\beta,\gamma\delta}\, \lambda^\alpha_a \lambda^\beta_a \lambda^\gamma_b \lambda^\delta_b \quad \text{no summation over }a,b\,,\\
(Q^{a}_{bc})_{\alpha\beta}&\Rightarrow Q^a_{bc}= (Q^{a}_{bc})_{\alpha\beta}\, \lambda^\alpha_a \lambda^\beta_a\quad \text{no summation over }a\,.
\end{align}
For instance, the correlator of two scalar operators $\mathcal O$ of the same dimension $\Delta$ with a conserved current of an integer spin-$s$ obeying \eqref{conservation-s} is
\begin{eqnarray} \label{bbJ-HS2}
\langle   O(X_1) O(X_2) J_{s}(X_3)    \rangle =C
{(\det |X_{12}|)^{-\frac{2-\Delta}2}  (\det |X_{13}|)^{-\frac 12} (\det |X_{23}|)^{-\frac 12}} (Q^3_{12})^s\,.
\end{eqnarray}
The current conservation condition leads to the same result as for the case of $s=1,2$, i.e. $k_1=k_2=\tfrac12$, which
means that the dimensions of the scalar operators are arbitrary.

However, if we consider a three-point function of a scalar operator and two conserved currents
\be
J_s(X)=J_{\alpha_1\ldots\alpha_{2s}}(X)\lambda^{\alpha_1}\cdots\lambda^{\alpha_{2s}}
\ee
of ranks $2s_1$ and $2s_2$ with $s\geq 1$, we will again find that, up to an  overall factor, all the free parameters in the correlator are fixed. For example,
\begin{equation} \label{bbJ-HS3}
\langle   J_3(X_1) J_1(X_2) O(X_3)    \rangle =C\frac{{(Q^1_{23})^3 Q^2_{13}-3 (Q^1_{23})^2 P_{12} }}{\Big(\det |X_{12}|\det |X_{13}|\det |X_{23}|\Big)^{1/2}}.
\end{equation}

From the discussion above, one can conclude that
 in order  to describe the $Sp(2n)$-invariant three-point functions, we can borrow the generating functions of 3-point correlators of free symmetric higher-spin fields in conventional conformal theories \cite{Giombi:2010vg,Colombo:2012jx,Didenko:2012tv,Gelfond:2013xt,Didenko:2013bj} simply because the $Sp(2n)$ group contains the corresponding conformal group $SO(2,D)$ as a subgroup, or, in other words, the correlators in the free CFTs can be covariantly embedded into the $Sp(2n)$ invariant correlators.
For example, a generating function of the three-point functions of currents built out of free scalars $b(X)$ is
\begin{equation}
    \langle J(X_1)J(X_2)J(X_3)\rangle=\frac{\cos(p_{12})\cos(p_{13})\cos(p_{23})\,\exp\left(\frac12[Q^1_{23}+Q^2_{13}+Q^3_{12}]\right)}{(\det |X_{12}|\det |X_{23}|\det |X_{13}|)^{1/2}}\,.
    \end{equation}
It contains the operators $J_{s}(X)$, $s=0,1,2,...$ and the correlator $\langle J_{s_1} J_{s_2} J_{s_3}\rangle$ is obtained as the coefficient in front of $(\lambda_1)^{2s_1}(\lambda_2)^{2s_2}(\lambda_3)^{2s_3}$.

The generating function obtained from the currents built out of the free fermions $f_\alpha(X)$ is
\begin{equation}
    \langle J(X_1)J(X_2)J(X_3)\rangle=\frac{\sin(p_{12})\sin(p_{13})\sin(p_{23})\,\exp\left(\frac12[Q^1_{23}+Q^2_{13}+Q^3_{12}]\right)}{(\det |X_{12}|\det |X_{23}|\det |X_{13}|)^{1/2}}\,.
\end{equation}
The generating function of multi-point correlators can be found in \cite{Didenko:2012tv,Gelfond:2013xt,Didenko:2013bj,Sleight:2016dba,Bonezzi:2017vha}.

The above expressions deal with the bosonic symmetric tensor currents of even rank. The generating function which produces 3-point correlators involving two fermionic currents of odd ranks is similar, see e.g. \cite{Maldacena:2011jn}.

As a further development of this subject, it would be of interest to carry out the study of other aspects of the Sp(2n)-invariant higher-spin  systems, in particular,  to explore their links to recent results on conformal higher-spin theories in $AdS_D$ backgrounds (see e.g. \cite{Tseytlin:2013jya, Metsaev:2014iwa, Metsaev:2014sfa,Nutma:2014pua,Beccaria:2015vaa}) and to  Sp(2n)-invariant unfolded higher-spin structures discussed in \cite{Sharapov:2017yde}.

\subsection{Breaking $Sp(2n)$ symmetry}

As it follows from the discussion above, in order to have an interacting  generalized conformal field
theory based on $Sp(2n)$
symmetry, one has to break this symmetry down to a subgroup.
Obviously, in order to still use
 $Sp(2n)$ symmetry as a symmetry of the theory, it should be broken
 spontaneously rather then explicitly.
 On the other hand, the question whether a symmetry is broken spontaneously or explicitly
 could be simpler to address if one had the corresponding Lagrangian, which would produce
 the field equations \p{b} and \p{f} (and/or their possible nonlinear or massive deformations). Unfortunately, such a Lagrangian is still lacking.

In this respect, let us mention that
 the issue of breaking $Sp(8)$ symmetry via current interactions in the unfolded formulation has been addressed
in \cite{Gelfond:2015poa}.
In particular, analyzing the system of equations
\be \label{G_V}
DC(x, \mu, {\overline \mu}) = F(\omega, J(x, \mu, {\overline \mu})),
\qquad D_2 J(x, \mu, {\overline \mu}) =0,
\ee
where $D=d+\omega$  is a spin connection, $J$ is a current which is billinear
in the higher-spin functional $C$ and $D_2$ is the corresponding kinetic operator
(see the discussion around the equation \p{Yrr}), the authors showed that the
$Sp(8)$ symmetry is broken to the four-dimensional conformal group $SO(2,4)$.

 In the hyperspace framework, one may try to approach this problem as follows.
 First, one should construct a nonlinear deformation of the equations \p{b} and \p{f}
 \begin{eqnarray}\label{b-d}
\partial_{\alpha \beta}
\partial_{\gamma \delta}\,b(X)-\partial_{\alpha \gamma}\partial_{\beta \delta}\,b(X)&=&F_{b}(b,f, A)\,,\\
\quad \partial_{\alpha \beta} f_\gamma(X)-\partial_{\alpha \gamma}
f_\beta(X)&=&F_{f}(b,f, A)\label{f-d}\,.
\end{eqnarray}
with some unknown functions  $F_{b}(b,f, A)$ and $F_{f}(b,f, A)$. It is natural to expect
that these functions depend also on higher-spin potentials $A$, in addition to the  higher-spin curvatures
contained in the hyperfields $b(X)$  and $f_\mu(X)$. Note that in the unfolded description of the $Sp(8)$-invariant system, higher-spin gauge potentials were introduced, at the linearized level, in \cite{Vasiliev:2007yc}. As a necessary step forward, one should understand whether and how the equations \eqref{b-d} may result from a (non-linear) generalization of the construction of  \cite{Vasiliev:2007yc}.

The right hand sides of the equations \eqref{b-d} should be chosen under the requirement that
the analysis of the equations \p{b-d} and \p{f-d}, similar to
the one carried out for the free equations in Subsection \ref{feI}
leads to a physically meaningful nonlinear equations in the $x$--space.
This is an interesting open problem for a future study.

\section{Conclusion}

The idea to formulate higher-spin theories in an extended (super)space, where  extra coordinates
generate higher spins (by analogy with the Kaluza-Klein theories where compact extra dimensions generate ``higher masses")
seems to be very attractive, especially taking into account a level of complexity of
higher-spin theories formulated in an ordinary space-time.

The underlying symmetry of this formulation is the $Sp(2n)$ group, which contains the corresponding $D$--dimensional
conformal group as a subgroup. This allows one to borrow, for the analysis of the $Sp(2n)$-invariant systems, an intuition and techniques from conventional Conformal Field Theories.

To summarize, the reviewed appraoch generalizes familiar concepts to higher-dimensional tensorial spaces and the correspondence
looks schematically as follows
\begin{itemize}
\item Space time-coordinates $x^m$ are extended to tensorial coordinates  $X^{\alpha \beta}$.

\item Cartan-Penrose relation $P_{A \dot A} = \lambda_A  \overline \lambda_{\dot A}$  gets extended to the hyperspace twistor-like relation $P_{\alpha \beta}=\lambda_\alpha \lambda_\beta$ which determines free dynamics of fields in the tensorial space with the momentum $P_{\alpha \beta}$ conjugate to $X^{\alpha \beta}$.

\item $AdS_D$ space is extended to the $Sp(n)$ group manifold.

\item Conformal scalar $\phi(x)$ and conformal spinor $\psi_\mu(x)$
become the `hyperscalar' $b(X)$ and the `hyperspinor' $f_\mu(X)$.

\item $D$-dimensional conformal group $SO(2,D)$ is extended to the $Sp(2n)$ group which underlies the Generalized Conformal Field Theory of the fields $b(X)$ and  $f_\mu(X)$.

\end{itemize}
We have shown that the hyperspace approach  describes (in $D=3,4,6$ and 10) free dynamics of an infinite set of massless conformal higher-spin fields in an elegant compact form. An important and non-trivial problem is to find a non-linear generalization of this formulation which would correspond to an interacting higher-spin theory.
This problem has been addressed by several authors. As we have seen, it is related to the necessity to break  the $Sp(2n)$ symmetry in an appropriate way.
Attempts to construct such a generalization in the
framework of hyperspace supergravity and a non-linear realization
of the $OSp(1|8)$ supergroup were undertaken, respectively, in
\cite{Bandos:2004nn} and \cite{Ivanov:2005ss}.
Obstacles encountered in these papers may be related to the fact that their constructions utilized only higher-spin field strengths but
did not include couplings to higher-spin gauge potentials, while the consistent formulation of nonlinear equations of massless
higher-spin fields contains both
\cite{Vasiliev:1995dn,Vasiliev:1999ba,Bekaert:2005vh}.
Therefore, 
to successfully address the problem of interactions it is important to incorporate higher-spin potentials in the hyperspace approach, e.g. by further elaborating on the construction of \cite{Vasiliev:2007yc}.

Another issue, which can be related to the previous one,
is a question of consistent breaking $Sp(2n)$
symmetry. The manifestation of this breaking was observed e.g. in higher-spin current interactions\cite{Gelfond:2015poa}.
As we have seen in Section \ref{sections6},
when considering generalized CFT based
on global $Sp(2n)$ invariance
(see \cite{Skvortsov:2016lbh}), the requirement of generalized current conservation
turns out to be too strong to allow for the basic hyperfields to have anomalous conformal dimensions and again points at the necessity to (spontaneously) break $Sp(2n)$ invariance.

Theories with spontaneously broken
$Sp(2n)$ symmetry  might be also useful for studying massive higher-spin fields in
hyperspaces. A consideration of theories with local $Sp(n)$ invariance i.e. some sort of generalized gravity
is yet another interesting and widely unexplored area.

Finally, let us mention that  field equations \eqref{b} and \eqref{f} for the fields in hyperspaces remind (a part of) weak section conditions of exceptional field theories (see \cite{Baguet:2015xha} for a review and references). This similarity
 can be relevant for higher-spin extensions of these theories, provided the section conditions can be properly relaxed (see e.g. \cite{Cederwall:2015jfa,Bandos:2016ppv} for a discussion of this point). It would be interesting to further elaborate on this issue, as a  connection to the $E_{11}$ framework \cite{West:2007mh}.

\vskip 0.5cm

\noindent {\bf Acknowledgments.} We are  grateful to I. Bandos, X. Bekaert,  J.A. de Azcarraga,
I. Florakis, J. Lukierski, P. Pasti, M. Plyushchay, E. Skvortsov and M. Tonin with whom we obtained the results
reviewed in this article.
We are thankful to P.Bouwknegt, O. Gelfond, A.R. Gover, S. Kuzenko, S. Sergeev, A.Tseytlin and especially to M. Vasiliev for many fruitful discussions.
M.T. is grateful  to the Department of Mathematics, the University of Auckland, New Zealand,  where part of this 
work was done.
Work of D.S. and M.T. was supported by the Australian Research Council grant DP160103633. 
Work of D.S. was also partially supported by the Russian Science
Foundation grant 14-42-00047 in association with Lebedev Physical Institute.

\renewcommand{\thesection}{A}

\renewcommand{\theequation}{A.\arabic{equation}}

\setcounter{equation}0
\appendix
\numberwithin{equation}{section}

\section{Conventions}\label{Appendix A}

The $\gamma$--matrices satisfy the following anti-commutation relations
\begin{equation}\label{1}
(\gamma^m)^\alpha{}_\delta (\gamma^n)^\delta{}_\beta
+
(\gamma^n)^\alpha{}_\delta (\gamma^m)^\delta{}_\beta
 = 2 \eta^{mn} \delta^\alpha_\beta \,,
\end{equation}
where $m$,  $n$ and other Latin letters are space-time vector indices,  and
$\alpha, \beta$ and other Greek letters label spinorial indices.
Throughout the paper  ``$(,)$" denotes symmetrization and ``$[,]$" denotes antisymmetrization with weight one.
The symplectic matrix $C^{\alpha \beta}=-C^{\beta\alpha}$  is used
to relate upper and lower spinorial indexes as follows
\begin{equation}\label{updown}
\mu^{\alpha}= C^{\alpha \beta} \mu_\beta, \quad \mu_{\alpha}= -C_{\alpha \beta} \mu^\beta,
\quad C^{\alpha \gamma}C_{\gamma \beta}=-\delta^\alpha_\beta \,.
\end{equation}
The differentiation by hypercoordinates $X^{\alpha \beta}$ is as follows
\begin{equation}\label{derivative-1}
\frac{ d X^{\alpha \beta}}{d X^{\gamma \delta}} \equiv \partial_{\alpha \beta}X^{\gamma \delta}=
\frac{1}{2}(\delta^\alpha_\gamma \delta^\beta_\delta +
\delta^\beta_\gamma \delta^\alpha_\delta) \,,
\end{equation}
\be
\partial_{\mu \nu} X^{-1}_{\alpha \beta}= - \frac{1}{2}(X^{-1}_{\mu \alpha} X^{-1}_{\nu \beta} + X^{-1}_{\mu \beta}X^{-1}_{\nu \alpha})
\ee
and
\be
\partial_{\mu \nu} ({\det X})= X^{-1}_{\mu \nu}({\det X})
\ee
where
\be
X^{-1}_{\mu \nu} X^{\nu \alpha}= \delta_\mu^\alpha.
\ee
Let us note that the product of an even number of $X^{\alpha \beta}$ matrices is antisymmetric  in spinorial indexes, whereas the product of an odd number of $X^{\alpha \beta}$ is a symmetric matrix.
For example,
\begin{equation}\label{PX}
X^{\alpha \gamma} X_{\gamma}{}^{\beta}=- X^{\beta \gamma}X_{\gamma}{}^\alpha,
\quad
X^{\alpha}{}_\gamma X^\gamma{}_\delta X^{\delta \beta}=
+X^\beta{}_\delta X^\delta{}_\gamma X^{\gamma \alpha}, \quad \textrm{etc}.
\end{equation}

\section{Derivation of the field equations on $Sp(n)$  } \label{derivationofeqs}
Let us   evaluate the operator $Y_{(\alpha}Y_{\beta)}$ in \eqref{ysp}:
\begin{align}
	\frac{1}{2}(Y_\alpha Y_\beta+Y_\beta Y_\alpha) \equiv Y_{(\alpha}Y_{\beta)} = (\tfrac{\xi}{8})^2\mu_\alpha \mu_\beta+\tfrac{i \xi}{8}\left(\mu_\alpha\tfrac{\partial}{\partial\mu^\beta}+\mu_\beta\tfrac{\partial}{\partial\mu^\alpha}\right)- \tfrac{\partial}{\partial\mu^\alpha}\tfrac{\partial}{\partial\mu^\beta}~.
\end{align}

\subsection{Fermionic equation}
Consider the equation \p{ysp}. Substituting 
into it the expansion \p{pol-1} one gets for the term
linear in $\mu^\alpha$
\begin{equation}
\nabla_{\alpha \beta} F_\gamma  (X) \, \mu^\gamma + \frac{\xi}{8}(C_{ \gamma \alpha} F_\beta (X) + C_{\gamma \beta} F_\alpha (X)
) \, \mu^\gamma=0
\end{equation}
The second term comes from $-\frac{i}{2}(Y_\alpha Y_\beta + Y_\alpha Y_\beta)$ acting on $F_\gamma \mu^\gamma$.
From this equation one gets \p{fsp}. 

\subsection{Bosonic equation}

The equation \p{ysp} to the  zeroth order in $\mu^\alpha$ becomes:
\begin{align}
		\nabla_{\alpha\beta}B (X) =i Y_{(\alpha}Y_{\beta)}\cdot\tfrac{1}{2}B_{\gamma\delta}(X)\mu^{\gamma}\mu^{\delta}~.
\end{align}
Obviously, only the double  $\mu$-derivative in $Y_{(\alpha}Y_{\beta)}$ will contribute to this order. Thus, we have:
\be
	\nabla_{\alpha\beta}B (X)=-i \tfrac{\partial}{\partial\mu^\alpha}\tfrac{\partial}{\partial\mu^\beta} \cdot \tfrac{1}{2}B_{(\gamma\delta)} (X) \mu^\gamma \mu^\delta  
\ee
Therefore,
\begin{align}\label{Bab}
	\nabla_{\alpha\beta}B (X) =-i  \, B_{(\alpha\beta)}(X)~,
\end{align}
Which indicates that all the higher order components in the expansion \eqref{pol-1} are expressed in terms of  $B(X)$ and $F_\alpha(X)$.

 To zeroth order in $\mu^\alpha$, we compute:
\begin{eqnarray}
	&& (\nabla_{\alpha\beta}-iY_{(\alpha}Y_{\beta)})(\nabla_{\gamma\delta}-iY_{(\gamma}Y_{\delta)})\cdot \\ \nonumber
&&	\cdot \Bigr[B(X)+\tfrac{1}{2}B_{\rho\sigma}(X)\mu^\rho \mu^\sigma+\tfrac{1}{4!}B_{\rho\sigma\tau\lambda}(X)\mu^\rho \mu^\sigma 
	 \mu^\tau \mu^\lambda+\ldots \Bigr] =0~.
\end{eqnarray}
\begin{align}\nonumber
	0=&\nabla_{\alpha\beta}\nabla_{\gamma\delta}B (X) +  (C_{\alpha\gamma}C_{\beta\delta}+C_{\beta\gamma}C_{\alpha\delta})B(X)
	+(\tfrac{\xi}{8})^2 B_{(\alpha\beta\gamma\delta)}(X) \\ \nonumber
	&+i(\tfrac{\xi}{8}) \Bigr[ C_{\alpha\gamma}B_{(\beta\delta)}(X)
	+C_{\alpha\delta}B_{(\beta\gamma)}(X)+C_{\beta\gamma}B_{(\alpha\delta)}(X)+C_{\beta\delta}B_{(\alpha\gamma)}(X)\Bigr] \\
	&+i \Bigr[ \nabla_{\gamma\delta}B_{(\alpha\beta)}(X)+\nabla_{\alpha\beta}B_{(\gamma\delta)}(X)\Bigr]~.
\end{align}
Now, using \eqref{Bab}, this becomes:
\begin{align}\nonumber
	0=&\nabla_{\alpha\beta}\nabla_{\gamma\delta}B(X) + (\tfrac{\xi}{8})^2 (C_{\alpha\gamma}C_{\beta\delta}+C_{\beta\gamma}C_{\alpha\delta})B(X)+ B_{(\alpha\beta\gamma\delta)} (X)\\ \nonumber
	&-\tfrac{\xi}{8} \Bigr[ C_{\alpha\gamma}\nabla_{\beta\delta}+C_{\alpha\delta}\nabla_{\beta\gamma}+C_{\beta\gamma}\nabla_{\alpha\delta}+C_{\beta\delta}\nabla_{\alpha\gamma}\Bigr]B(X) \\
	&- \Bigr[ \nabla_{\gamma\delta}\nabla_{\alpha\beta}+\nabla_{\alpha\beta}\nabla_{\gamma\delta}\Bigr]B(X)~.
\end{align}
Using the  algebra  \eqref{algebra} for the covariant derivatives
$\nabla_{\alpha\beta}$, we can write:
\begin{align}\label{intermedio}
	\nabla_{\gamma\delta}\nabla_{\alpha\beta}B(X) =&  (\tfrac{\xi}{8})^2 (C_{\alpha\gamma}C_{\beta\delta}+C_{\beta\gamma}C_{\alpha\delta})B(X)+ B_{(\alpha\beta\gamma\delta)}(X) -\tfrac{1}{2} [\nabla_{\alpha\beta},\nabla_{\gamma\delta}] B (X) ~.
\end{align}
From this equation, we obtain the bosonic equation \p{bsp}. Let us note that 
 exchange of indexes as $\alpha\leftrightarrow\gamma$ and $\beta\leftrightarrow\delta$ :
\begin{align} \label{B11}
	\nabla_{\alpha\beta}\nabla_{\gamma\delta}B (X) =&  (\tfrac{\xi}{8})^2 (C_{\alpha\gamma}C_{\beta\delta}+C_{\beta\gamma}C_{\alpha\delta})B (X)+ B_{(\alpha\beta\gamma\delta)} (X) +\tfrac{1}{2} [\nabla_{\alpha\beta},\nabla_{\gamma\delta}] B (X)~.
\end{align}
and subtraction of \p{intermedio} and \p{B11} leads to an identity.

\section{~Some ~identities ~for ~supercoordinates ~on $OSp(1|n)$ ~group manifold}\label{CC}
The supercoordinates on $OSp(1|n)$ group manifold obey some useful relations in particular
\begin{equation}\label{Theta1}
\theta^\alpha{\mathcal G}_{\alpha}{}^\beta=\Theta^\beta P(\Theta^2), \qquad \theta^\alpha=\Theta^\beta {\mathcal G}_{\beta}^{-1\alpha}P(\Theta^2)\,,
\end{equation}
\be\label{2-1}
Q_\beta\Theta^\alpha=P^{-1}(\Theta^2)\left(G_\beta{}^\alpha +\frac{i\xi}8
\Theta_\beta \Theta^\alpha +
\frac{i\xi}8 G_\beta{}^\sigma
\Theta_\sigma \Theta^\alpha
+
\left(\frac{i\xi}8\right)^2 \Theta^2 \Theta_\beta\Theta^\alpha\right)\,,
\ee
\be\label{3-1}
(Q_\beta\Theta^\alpha)\Theta_\alpha=P(\Theta^2)\left(G_\beta{}^\sigma
 +\frac{i\xi}{8}\Theta_\beta\Theta^\sigma\right)\Theta_\sigma,
\ee
\be \label{4-1}
\partial_{\alpha\beta}\Theta^\gamma =\frac\xi 4\Theta_{(\alpha}G_{\beta)}{}^\delta(\delta_\delta^\gamma+\frac{i\xi}8\Theta_\delta\Theta^\gamma)\,,
\ee
\be\label{5-1}
D_\beta{\mathcal G}_\alpha{}^\gamma =\frac{i\xi}4P(\Theta^2)\,(\Theta_\alpha-2G_\alpha{}^\rho\Theta_\rho){\mathcal G}_\beta{}^\gamma
\ee
\be\label{6-1}
\partial_{\alpha\beta}{\mathcal G}_\gamma{}^\delta=\frac \xi 4{\mathcal G}_{\gamma(\alpha}\,{\mathcal G}_{\beta)}{}^\delta\,,
\ee
and
\be \label{7-1}
Q_\alpha {\mathcal G}_{\mu \nu} =  -\frac{i\xi}{4} P(\Theta^2) \Theta_\nu {\mathcal G}_{\mu \alpha}\,,
\ee

\if{}
\bibliographystyle{abe}
\bibliography{references}
\fi

\providecommand{\href}[2]{#2}\begingroup\raggedright\endgroup

\end{document}